\documentclass[pra,twocolumn,showpacs,superscriptaddress,floatfix]{revtex4}

\let\a=\alpha \let\b=\beta  \let\g=\gamma  \let\d=\delta \let\e=\varepsilon
\let\z=\zeta  \let\h=\eta   \let\th=\theta \let\k=\kappa \let\l=\lambda
\let\m=\mu    \let\n=\nu    \let\x=\xi         \let\r=\rho
\let\s=\sigma \let\t=\tau   \let\f=\varphi 
\let\ch=\chi  \let\ps=\psi   \let\o=\omega
\let\G=\Gamma \let\D=\Delta  \let\Th=\Theta\let\L=\Lambda \let\X=\Xi
    \let\Si=\Sigma \let\F=\Phi    
\let\O=\Omega 

\font\tenmib=cmmib10\font\sevenmib=cmmib7\font\fivemib=cmmib5%
\mathchardef\Ba   = "050B  
\mathchardef\Bb   = "050C  
\mathchardef\Bg   = "050D  
\mathchardef\Bd   = "050E  
\mathchardef\Be   = "0522  
\mathchardef\Bee  = "050F  
\mathchardef\Bz   = "0510  
\mathchardef\Bh   = "0511  
\mathchardef\Bthh = "0512  
\mathchardef\Bth  = "0523  
\mathchardef\Bi   = "0513  
\mathchardef\Bk   = "0514  
\mathchardef\Bl   = "0515  
\mathchardef\Bm   = "0516  
\mathchardef\Bn   = "0517  
\mathchardef\Bx   = "0518  
\mathchardef\Bom  = "0530  
\mathchardef\Bp   = "0519  
\mathchardef\Br   = "0525  
\mathchardef\Bro  = "051A  
\mathchardef\Bs   = "051B  
\mathchardef\Bsi  = "0526  
\mathchardef\Bt   = "051C  
\mathchardef\Bu   = "051D  
\mathchardef\Bf   = "0527  
\mathchardef\Bff  = "051E  
\mathchardef\Bch  = "051F  
\mathchardef\Bps  = "0520  
\mathchardef\Bo   = "0521  
\mathchardef\Bome = "0524  
\mathchardef\BG   = "0500  
\mathchardef\BD   = "0501  
\mathchardef\BTh  = "0502  
\mathchardef\BL   = "0503  
\mathchardef\BX   = "0504  
\mathchardef\BP   = "0505  
\mathchardef\BS   = "0506  
\mathchardef\BU   = "0507  
\mathchardef\BF   = "0508  
\mathchardef\BPs  = "0509  
\mathchardef\BO   = "050A  
\mathchardef\BDpr = "0540  
\mathchardef\Bstl = "053F  

\newdimen\xshift \newdimen\xwidth \newdimen\yshift \newdimen\ywidth

\def\ins#1#2#3{\vbox to0pt{\kern-#2pt\hbox{\kern#1pt #3}\vss}\nointerlineskip}

\def\eqfig#1#2#3#4#5{
\par\xwidth=#1pt \xshift=\hsize \advance\xshift
by-\xwidth \divide\xshift by 2
\yshift=#2pt \divide\yshift by 2
{\hglue\xshift \vbox to #2pt{\vfil
#3 \includegraphics{#4.eps}
}\hfill\raise\yshift\hbox{#5}}}

\def\V#1{{\bf #1}}
\def\lis#1{{\overline#1}}

\font\titolo=cmbx12
\def\tende#1{\,\vtop{\ialign{##\crcr\rightarrowfill\crcr
 \noalign{\kern-1pt\nointerlineskip} \hskip3.pt${\scriptstyle
 #1}$\hskip3.pt\crcr}}\,}
\def\AA{{\cal A}}\def\XX{{\cal X}}
\def\EE{{\cal E}}\def\BB{{\cal B}}
\def\HH{{\cal H}}\def\NN{{\cal N}}\def\CC{{\cal C}}

\font\msytw=msbm10
\font\msytww=msbm8 
\def\RRR{\hbox{\msytw R}}\def\rrr{\hbox{\msytww R}}

\def\ZZZ{\hbox{\msytw Z}}

\def\defi{\,{\buildrel def\over=}\,}
\let\wt=\widetilde

\def\be{\begin{equation}}\def\ee{\end{equation}}
\renewcommand{\theequation}{\arabic{section}.\arabic{equation}}
\let\0=\noindent
\def\*{\vskip2mm}

\def\Eq#1{{\label{#1}}}%
\def\equ#1{(\ref{#1})}
\def\eq#1{(\eqlab{#1})}%
\usepackage{eqalignno}
\usepackage{fancyhdr}
\def\iniz{\setcounter{equation}{0}}

\makeatletter

\newcommand{\Rmnum}[1]{\expandafter\@slowromancap\romannumeral #1@}
\makeatother
\begin{document}

\centerline{\titolo
Thermodynamic limit of isoenergetic}
\centerline{\titolo and Hamiltonian Thermostats}
{\vskip3mm}

\centerline{G. Gallavotti${}^*$ and E. Presutti${}^@$ }
\centerline{${}^*$ Fisica-INFN Roma1 and Rutgers U.}
\centerline{${}^@$ Matematica Roma2}
\centerline{\today}

{\vskip3mm}
\noindent {\bf Abstract}: {\it The relation between
 isoenergetic and Hamiltonian thermostats is
 studied and their equivalence in the thermodynamic limit is
 proved in space dimension $d=1,2$.}  {\vskip3mm}

\0{\bf Contents}{\small
\halign{# &\kern3mm#\hfill&\kern3mm\hfill#\cr
{\Rmnum 1}& Thermostats&\pageref{sec1}
\cr
{\Rmnum 2}&  Notations and sizes&\pageref{sec2}
\cr
{\Rmnum 3}&  Equivalence: isoenergetic versus
Hamiltonian&\pageref{sec3}
\cr
{\Rmnum 4}&  Free thermostats&\pageref{sec4}
\cr
{\Rmnum 5}&  Kinematics&\pageref{sec5}
\cr
{\Rmnum 6}&  Energy bound&\pageref{sec6}
\cr
{\Rmnum 7}&  Entropy bound and phase space contraction&\pageref{sec7}
\cr
{\Rmnum 8}&  Infinite volume Hamiltonian dynamics&\pageref{sec8}
\cr
{\Rmnum 9}&  Infinite volume thermostatted dynamics&\pageref{sec9}
\cr
{\rm X}&  Conclusions&\pageref{sec10}
\cr
{\rm XI}&  Appendices&\pageref{sec11}
\cr}}

\def\SEC{Thermostats}
\section{Thermostats}\label{sec1}
\makeatletter
\def\iniz{\setcounter{equation}{0}
\ifodd\thepage 
{\rhead{\thepage}\lhead{{{{\small\bf\thesection:}\ \SEC}}}}
\else
\lhead{\thepage}\rhead{{\small\bf\thesection:}\ \SEC}
\fi}
\makeatother
\iniz
In a recent paper \cite{Ga008d} equivalence between isokinetic and
Hamiltonian thermostats has been discussed heuristically, leaving
aside several difficulties on the understanding of the classical
dynamics of systems of infinitely many particles. Understanding it is,
however, a necessary prerequisite, because strict equivalence can be
expected to hold only in the thermodynamic limit. In this paper we
proceed along the same lines, comparing the isoenergetic and the
Hamiltonian thermostats, and study the conjectures corresponding to
the ones formulated in \cite{Ga008d} for isokinetic thermostats,
obtaining a complete proof of equivalence in $1$ and $2$--dimensional
systems.

Here the class of models to which our main result applies is described
in detail. The main result is informally quoted at the end of
Sec.\ref{sec1} after discussing the physics and the equations of
motion of the models; a precise statement will be theorem 1 in
Sec.\ref{sec3} and it will rely on a property that we shall call {\it
local dynamics}: the proof is achieved by showing that in the models
considered the local dynamics property holds as a consequence of the
theorems 2-9, each of which is interesting on its own right, discussed
in the sections following Sec.\ref{sec3}.  \vskip2mm

A classical model for nonequilibrium statistical mechanics, {\it e.g.}
see \cite{FV63}, is a {\it test system} in a container $\O_0$, and one
or more containers $\O_j$ adjacent to it and enclosing the {\it
interaction systems}.

\vskip2mm

\eqfig{200}{86}{%
\ins{40}{20}{%
$x=(\V X_0,\dot{\V X}_0,\V X_1,\dot{\V X}_1,\ldots,\V X_\n,\dot{\V X}_\n)$}
}{fig1}{}
\vskip-2mm

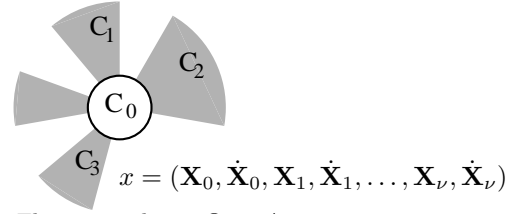
\noindent{Fig.1: \small\it The $1+\n$ boxes $\O_j\cap\L,\, j=0,\ldots,\n$,
  are marked $\CC_0,\CC_1,\ldots,\CC_\n$ and contain $N_0,N_1,\ldots, N_\n$
  particles with positions and velocities denoted $\V
  X_0,\V X_1,\ldots,\V X_\n$, and $\dot{\V X}_0,\dot{\V X}_1,\ldots,$
  $\dot{\V X}_\n$, respecti\-vely.
}  {\vskip2mm}

To fix the ideas the geometry that will be considered can be imagined
(see Fig.1, keeping in mind that it is just an example for convenience
of exposition and which could be widely changed) as follows: \vskip2mm

\noindent(1) The {\it test system} consists of particles enclosed in a
sphere $\O_0=\Si(D_0)$ of radius $D_0$ centered at the origin.
\vskip2mm

\noindent(2) The {\it interaction systems} consist of particles
enclosed in regions $\O_j$ which are disjoint sectors in $\RRR^d$,
{{\it i.e.}}  disjoint semiinfinite ``spherically truncated'' cones
adjacent to $\O_0$, of opening angle $\o_j$ and axis $\V k_j$:
$\O_j =\{\x\in \RRR^d, |\x|>D_0, \x\cdot\V
k_j<|\x|\,\o_j\}, \,j=1,\ldots,\n$ .
\vskip2mm

The initial configurations $x$ of positions and velocities will be
supposed to contain finitely many particles in each unit cube. Thus
the {\it test system} will consist of {\it finitely many
particles}, while the {\it interaction systems are infinitely
exten\-ded}.

The motion starting from $x$ must be defined by first {\it
regularizing} the equations of motion (which are infinitely many and
therefore a ``solution'' has to be shown to exist). The regularization
considered here will be that only the (finitely many) particles of the
initial data $x$ inside an artificial finite ball $\L=\Si(r)$ of
radius $r>D_0$ will be supposed moving.

{\it I.e.} for the same initial data $x$ only the particles in
$\O_0,\O_1\cap\L,\ldots,\O_\n\cap \L$ will move and kept inside $\L$ by
an {\it elastic reflection} boundary condition at the boundary of
$\L$, while they will never reach the boundaries of the $\O_j$'s
because of the action of a force, modeling the walls of the $\O_j$ and
diverging near them.

The particles of $x$ located outside the container $\O_0\cup\cup_{j>0}
(\O_j\cap\L)$ are imagined immobile in the initial positions and
influence the moving particles only through the force that the ones of
them close enough to the boundary of $\L$ exercise on the particles
inside $\L$.

In the ``{\it thermodynamic limit}'', which will be of central
 interest here, the ball $\L$ grows to $\infty$ and the particles that
 eventually become internal to $\L$ start moving: in other words we
 approximate the infinite volume dynamics with a finite volume one,
 called {\it $\L$--regularized}, and then take an infinite volume
 limit.

A configuration $x$ will be imagined to consist of
a configuration $(\V X_0,\dot {\V X}_0)$ with ${\bf X}_0$
contained in the sphere $\Si(D_0)$, delimiting the
container $\O_0$ of the test systems, and by $n$ configurations $(\V
X_j,\dot{\V X}_j)$ with $\V X_j\subset \O_j\cap \RRR^d)$, $j=1,\ldots,\n$:

{\vskip3mm} \noindent {\bf Phase space:} {\it Phase space $\cal H$ is
  the collection of locally finite particle configurations $x=(\ldots,
  q_i,\dot q_i,\ldots)_{i=1}^\infty$
\be x=(\V X_0,\dot{\V X}_0, \V X_1,\dot{\V X}_1, \ldots, \V
X_\n,\dot{\V X}_\n)\defi(\V X,\dot{\V X})
\Eq{e1.1}\ee
with $\V X_j\subset\O_j$ and
$\dot q_i\in \RRR^d$: in every ball $\Si(r')$ of radius $r'$ and
center at the origin $O$, fall a finite number of points of $\V X$. If
$\L=\O\cap \Si(r)$ and $x\in\HH$ we shall denote $x_\L$ the positions
and velocities of the particles of $x$ located in $\L$ together with
the positions of the particles of $x$ outside $\L$.}

{\vskip2mm} The particles of $x$ located outside $\L$ will be
regarded as immobile.  The particles are supposed to interact with
each other via a
potential $\f$ and with the walls with a potential $\ps$:

{\vskip2mm}

\noindent {\bf Interaction:} {\it Interparticle interaction will be
through a pair potential $\f$ with finite range $r_\f$ smooth,
decreasing and positive at the origin. The walls of the containers
$\O_j$ are represented by a smoooth decreasing potential $\ps\ge0$ of
range $r_\ps\ll r_\f$ and diverging as an inverse power of the
distance to the walls.}

{\vskip2mm} Hence the potential $\f$ is {\it superstable} in the sense
of \cite{Ru970}: a property that will play an important role in the
following.
The value of the potential $\f$ at midrange will be denoted $\lis\f$
and $0<\lis\f<\f_0\defi \f(0)$; the wall potential at a point $q$ at
distance $r$ from a wall will be supposed to be given by

\be\ps(q)=\big(\frac{r_\ps}{2r}\big)^{\a}\f_0 ,\qquad r\le
\frac{r_\ps}2
\Eq{e1.2}\ee
with $\a>0$ and $r$ equal to the distance of $q$ to the wall; for
larger $r$ it continues, smoothly decreasing, reaching the value $0$
at $r=r_\ps$.  The choice of $\ps$ as proportional to $\f_0$ limits
the number of dimensional parameters, but it could be made
general. The restriction $r_\ps\ll r_\f$ is required to facilitate the
interaction between particles in $\O_0$ and particles in $\cup_{j>0}
\O_j$.

The particles in $\O_0$ are supposed to interact with all the others
but the particles in $\O_j$ interact only with the ones in
$\O_j\cup\O_0$: {\it the test system in $\O_0$ interacts with all
thermostats but each thermostat interacts only with the system}, see
Fig.1.

The equations of the $\L$--regularized motion (see Fig.1), aside from
the reflecting boundary condition on the artificial boundary of $\L$,
concern only the particles in $\O_0\cup\cup_{j>0} (\O_j\cap\L)$ and
will be

$$\eqalignno{ m\ddot{\V X}_{0i}=&-\partial_i U_0(\V X_0)-\sum_{j>0}
\partial_i U_{0,j}(\V X_0,\V X_j)+\BF_i(\V X_0)\cr m\ddot{\V
X}_{ji}=&-\partial_i U_j(\V X_j)- \partial_i U_{0,j}(\V X_0,\V
X_j)-a\,\a_j \V{{\dot X}}_{ji} &\eq{e1.3}\cr}$$
where (1) the parameter
$a$ will be $a=1$ or $a=0$ depending on the model considered;
\*
\0(2) the potential energies $U_j(\V X_j), \,j\ge0$
and, respectively, $U_{0,j}(\V X_0,\V X_j)$ denote the internal
energies of the various systems and the potential energy of
interaction between the system and the thermostats; hence
for $\V X_j\subset\O_j\cap\L$ the $U_j$'s are:

\be \eqalign{
U_j(\V X_j)=&\sum_{q\in\V X_j}\ps(q)+\sum_{q,q'\in\V X_j, q\in\L}\f(q-q')\cr
U_{0,j}(\V X_0,\V X_j)=&\sum_{q\in\V X_0,q'\in\V X_j}\f(q-q');\cr}
\Eq{e1.4}\ee
\*
\0(3) the first label in Eq.\equ{e1.3}, $j=0$ or $j=1,\ldots,\n$
respectively, refers to the test system or to a thermostat, while the
second indicates the derivatives with respect to the coordinates of
the points in the corresponding container. Hence the labels $i$ in the
subscripts $(j,i)$ have $N_j$ values and each $i$ corresponds (to
simplify the notations) to $d$ components;
\*
\0(4) the multipliers $\a_j$
are, for $j=1,\ldots,\n$,

\be \eqalign{
\a_j{\,{\buildrel def\over=}\,}&\frac{Q_j}{d\,N_j
 k_B T_j(x)/m},\qquad{\rm with}\cr Q_j{\,{\buildrel def\over=}\,}&
-\dot {\V X}_j\cdot \partial_j
U_{0,j}(\V X_0,\V X_j),\cr} \Eq{e1.5} \ee
where $\frac{d}2\, N_j\,k_B\, T_j(x){\buildrel def\over=}
K_{j,\L}(\dot{\V X}_j)\defi \frac{m}2\dot{\V X}_j^2$and $\a_j$ are
chosen so that $K_{j,\L}(\dot{\V X}_j)+U_{j,\L}(\V X_j)=E_{j,\L}$ are
{\it exact constants of motion if $a=1$}: the subscript $\L$ will be
omitted unless really necessary. A more general model to which the
analysis that follows also applies is in \cite{Ga06c}. 
\*
\0(5) The
forces $\BF(\V X_0)$ are, positional, {\it nonconservative}, smooth
``stirring for\-ces'', possibly absent.  
\*
\0(6)\label{t} In the case of $\L$-regularized thermostatted dynamics
we shall consider only initial data $x$ for which the kinetic energies
$K_{j,\L}(\dot{\V X}_j)$ of the particles in the $\O_j\cap\L$'s are
$>0$ for all large enough $\L$. Then the time evolution is well
defined for $t<t_\L(x)$ where $t_\L(x)$ is defined as the maximum time
for which the kinetic energies remain positive (hence the equations of
motion remain well defined because the denominators in the $\a_j$ stay
positive). It will be important to remark that if $t_\L(x)<+\infty$
the moving particles positions and velocities have a limit even as
$t\to t_\L(x)$ because the accelerations $\a_j \dot q_{ji}$ remain
bounded even though $\a_j\to\infty$ (by the Schwartz inequality a
bound on $\a_j \dot q_{ji}$ could be $N_ {\L}^2\,|\max |\partial\f|$ if
$N_\L$ is the number of particles in $\L$).  \vskip2mm

The equations of motion with $a=1$ will be called $\L$--regularized
{\it isoenergetically thermostatted} because the energies
$E_j=K_j+U_j$ stay exactly constant for $j>0$ and equal to their
initial values $E_j$.  The equations with $a=0$ in Eq.\equ{e1.3} will
be considered together with the above and called the $\L$--regularized
{\it Hamiltonian equations}.

The qualifier ``Hamiltonian'' refers to case $a=0$ in which {\it no
dissipation} occurs even though, strictly speaking, the equations,
unless $\BF=0$, are still not Hamiltonian (in spite of $\a_j=0$).
\vskip2mm

{\it Remark that $Q_j$ is the work done, per unit time, by the test
system on the particles in the $j$-th thermostat.}  \vskip2mm

\vskip2mm

The {\it essential physical requirement} that the thermo\-stats should
have a well defined temperature and density is satisfied by an
appropriate selection of the inital conditions. The guiding idea is
that the thermostats should be so large that the energy that the test
system transfers to them, per unit time in the form of work $Q_j$, is
acquired without changing, not at least in the thermodynamic limit,
the average values of the densities and kinetic energies ({\it i.e.}
temperatures) of the thermostats in any finite
observation time $\Th>0$..

To impose the latter requirement, in the thermodynamic limit, the
values $N_j, E_j$ will be such that $\frac{N_j}{|\O_j\cap\L|}$
$\tende{\L\to\infty} \d_j$ and, for $j>0$,
$\frac{E_j}{|\O_j\cap\L|}\tende{\L\to\infty} e_j$: with $\d_j,e_j >0$
fixed in a sense that is specified by a choice of the initial data
that will be studied, and whose physical meaning is that of imposing
the values of density and temperature in the termostats, for $j>0$.

{\vskip3mm} \noindent {\bf Initial data:} {\it The probability
  distribution $\m_0$ for the random choice of initial data will be,
  if $dx{\,{\buildrel def\over=}\,}\prod_{j=0}^\n\frac{ d\V
  X_j\,d\dot{\V X}_j}{N_j!}$, the limit as $\L_0\to\infty$ of the
  finite volume grand canonical distributions on ${\cal H}$
$$\eqalignno{
&\m_{0,\L_0}(dx)=const\,\, e^{-H_{0,\L_0}(x)}
\,dx,\qquad{\rm with}&{\rm\eq{e1.6}}\cr
&H_{0,\L_0}(x)\defi \sum_{j=0}^\n \b_j (K_{j,\L_0}
(x)-\l_j N_{j,\L_0}+ U_{j,\L_0}(x))\cr
&\b_j{\,{\buildrel def\over=}\,} \frac1{k_BT_j}>0,\,\l_j\in \RRR,
\cr}$$
}

\noindent{\it Remarks:} (a) The values $ \b_0=\frac1{k_BT_0}>0,\l_0\in\RRR$,
are also fixed, although they bear no particular physical meaning because the
test system is kept finite.
\\
(b) Here $\Bl=(\l_0,\l_1,\ldots\l_\n)$ and ${\bf T}=(T_0,T_1,\ldots, T_\n)$
are fixed {\it chemical potentials} and {\it temperatures}, and $\L_0$
is a ball centered at the origin and of radius $r_0$.
\\
(c) The distribution $\m_0$ is a Gibbs distribution
obtained by taking the ``thermodynamic limit'' $\L_0\to\infty$.
Notice that $\m_0$ is a product of independent Gibbs distributions
because $H_0$ does not contain the interaction potentials $U_{0,j}$.
\\ (d) $\L_0$ should not be confused with the regularization sphere
$\L$: it is introduced here and made, right away, $\infty$ only to
define $\m_0$.
\\ (e) The theory of the thermodynamic limit implies the existence of
the limit distribution $\m_0$, either at low density and high
temperature or on subsequences, \cite{Ga00}. In the second case
(occurring when there are phase transitions at the chosen values of
the thermostats parameters) boundary conditions have to be imposed
that imply that the thermostats are in a pure phase: for simplicity
such exceptional cases will not be considered; this will be referred
to as a ``no-phase transitions'' restriction.

\vskip2mm

\0{\bf Main result: }{\it In the thermodynamic limit, the thermostatted
evolution, with\-in any prefixed time interval $[0,\Th]$, becomes the
Hamiltonian evolution at least on a set of configurations which have
probability $1$ with respect to the initial distribution $\m_0$, in
spite of the non stationarity of the latter.}  \vskip2mm

\def\SEC{Notations and Sizes}
\section{Notations and Sizes}\label{sec2}
\iniz

The initial data will be naturally chosen at random with
respect to $\m_0$.  Let the ``pressure'' in the $j$-th thermostat be
defined by $p_j(\b,\l;\L_0) {\buildrel def\over=}
\frac1{\b\,|\O_j\cap\L_0|}$ $\log Z_{j,\L_0}(\b,\l)$ with

\be \eqalign{
Z_{\L_0}(\b,\l)=&\sum_{N=0}^\infty \int \frac{dx_N}{N!}\cr
&\cdot  e^{-\b(-\l
  N+K_{j}(x_N)+U_{j}(x_N))}\cr}
\Eq{e2.1}\ee
where the integration is over positions and velocities of the particles
in $\L_0\cap\O_j$. Defining $p(\b,\l)$ as the thermodynamic limit,
$\L_0\to\infty$, of $p_j(\b,\l;\L_0)$ we shall say that the
thermostats have densities $\d_j$, temperatures $T_j$, energy
densities $e_j$ and potential energy densities $u_j$, for $j>0$, given
by equilibrium themodynamics, {\it i.e.}:

$$\eqalignno{
\d_j=&-\frac{\partial p( \b_j,\l_j)}{\partial\l_j},\quad
k_BT_j= \b_j^{-1}&\eq{e2.2}\cr
e_j=&-\frac{\partial \b_jp(\b_j,\l_j)}{\partial \b_j}-\l_j\d_j,
\qquad u_j=e_j-\frac{d}2 \d_j\b_j^{-1}\cr}$$
which are the relations linking density $\d_j$, temperature
$T_j=(k_B\b_j)^{-1}$, energy density $e_j$ and potential energy
density $u_j$ in a grand canonical ensemble.

In general the $\L$--regularized time evolution changes the measure of a
volume element in phase space by an amount related to (but different
from) the variation of the Liouville volume. The variation per unit
time and unit mass of a volume element, measured via $\m_0$
in the sector of phase space containing $N_j>0$ particles in
$\O_j\cap\L,\, j=0,1,\ldots,\n$,  can be computed  and
is, under the $\L$--regularized dynamics,

\be \s(x)=\sum_{j>0}\frac{Q_j}{k_B T_j(x)}\,{(1-\frac1{d\,N_j})}+
\b_0 (\dot K_0+\dot
U_0)\Eq{e2.3}) \ee
as it follows by adding the time derivative of $\b_0(K_0+U_0)$ to the
divergence of Eq.\equ{e1.3} (regarded as a first order equation for
the $q$'s and $\dot q$'s) using the expression in Eq.\equ{e1.5} for
$\a_j$.  \vskip2mm

\noindent {\it Remarks:} (1) The dynamics given by the
Eq.\equ{e1.3} or by the same equations with $\a_j\equiv0$ are different.

\noindent (2) The relation $\b_0\,(\dot K_0+\dot U_0)=\b_0\,(\BF\cdot \dot{\V
  X}_0-\sum_{j>0}(\dot U_{0j}-Q_j))$ is useful in studying Onsager
reciprocity and Green-Kubo formulae, \cite{Ga008c}.

\noindent (3) It is also interesting to consider {\it isokinetic}
thermostats: the multipliers $\a_j$ are then so defined that $K_j$ is
an exact constant of motion: calling its value $\frac32 N_jk_B T_j$
the multiplier $\a_j$ becomes

\be \a_j{\,{\buildrel def\over=}\,}\frac{Q_j- \dot U_j}{d\,N_j
 k_B T_j/m}, \Eq{e2.4} \ee
with $Q_j$ defined as in Eq.\equ{e1.5}. They have been studied
heuristically, from the present point of view, in \cite{Ga008d}.
\vskip2mm

Choose initial data with the distribution $\m_{0} $ and let $x\to
S^{(\L,a)}_tx,\, a=0,1$, be the solution of the $\L$--regularized
equations of motion with $\a_j=0$ ($a=0$, ``Hamiltonian thermostats'')
or a;ternatively $\a_j$ given by Eq.\equ{e1.5} ($a=1$, ``isoenergetic
thermostats''), {\it assuming $t<t_\L(x)$}. 

Let $S^{(a)}_tx$ be the infinite volume dynamics $\lim_{\L\to\infty}$ $
S_t^{(\L,a)}x$, $a=0,1$, {\it provided the limit exists}. Let

\be\eqalign{
x^{(\L,1)}(t)\defi& S^{(\L,1)}_t x,\qquad
x^{(\L,0)}(t)\defi S^{(\L,0)}_t x,\cr
x^{(0)}(t)\defi& S^{(0)}_t x.\cr}\Eq{e2.5}\ee
In the Hamiltonian case the existence of a solution to the equations
of motion poses a problem only if we wish to study the $\L\to\infty$
limit, {{\it i.e.\ }} in the case in which the thermostats are
infinite. For $\L$ finite $S^{(\L,0)}_tx$ is well defined with
$\m_0$-probability $1$ as in \cite{MPPP976}.

In the thermostatted case the kinetic energy appearing in the
denominator of $\a_j$, see Eq.\equ{e1.5}, can be supposed to be $>0$
with $\m_0$--probability $1$. However it can become $0$ at some later
time $t_{\L}(x)$ (see item (6), p.\pageref{t}, and the example at the
end of Sec.\ref{sec3}). In the course of the analysis it will be
proved that with $\m_0$--probability $1$ it is
$t_{\L}(x)\tende{\L\to\infty}\infty$; therefore $S^{(\L,1)}_tx$ is
eventually well defined.

We shall denote $(\V X_j^{(\L,a)}(t),\dot{\V X}_j^{(\L,a)}(t))$ or
$(S_t ^{(\L,a)}x)_j$ or $x^{(\L,a)}_j(t)$ the positions and velocities
of the particles of $S^{(\L,a)}_t x$ in $\O_j$. And by
$x^{(\L,a)}_{ji}(t)$ the pairs of positions and velocities
$(q^{(\L,a)}_i(t),\dot q_i^{(\L,a)}(t))$ with $q_i\in\O_j$.

Then a particle with coordinates $(q_i,\dot q_i)$ at $t=0$ in, {\it
say}, the $j$-th container evolves, see Eq.\equ{e1.3}, as

\kern-6mm
\be\eqalign{
q_i(t)=&q_i(0)+\int_0^t\dot q_i(t')\,dt'\cr
\dot q_i(t)=&e^{-\int_{0}^ta\,\a_j(t')dt'}\dot q_i(0)\cr
&+\int_{0}^t
e^{-\int_{t''}^t a\,\a_j(t')dt'} \frac{F_i(t'')}m \,dt''\cr}\Eq{e2.6}\ee
where $F_i(t)=-\partial_{q_i} \big(U_j(\V X_j(t))+U_{j,0}(\V X_0(t),{\V
X}_j(t))\big)+\d_{j0}\F_i(\V X_0(t))$ and $\V X_j(t)$ denotes $\V
X_j^{(\L,a)}(t)$ or $\V X^{(0)}_j(t)$.

The first difficulty with infinite dynamics is to show that the number
of particles, and their speeds, in a finite region $\L$ remains finite and
bounded only in terms of the region diameter $r$ (and of the initial data):
for all times or, at least, for any prefixed time interval.

It is convenient to work with dimensionless quantities: therefore
suitable choices of the units will be made. If $\Th$ is a prefixed
time which is the maximum time that will be considered, then

\be\eqalign{
&\f_0\,: \, \hbox{(energy scale)},\
r_\f\,: \, \hbox{(length scale)},\cr
&\Th \,: \, \hbox{(time scale)},\
v_1=\sqrt{\frac{2\f(0)}m}\,
\ \hbox{(velocity scale)}\cr}\Eq{e2.7}\ee
are natural units for measuring energy, length, time, velocity,
respectively.

It will be necessary to estimate quantitatively the size of various
kinds of energies of the particles, of a configuration $x$, which are
localized in a region $\D$. Therefore introduce, for any region $\D$,
the following dimensionless quantities:

\be\eqalign{
(a)\ &N_\D(x),N_{j,\D}(x)\ \hbox{the number of particles of $x$}\cr
& \hbox{located in $\D$ or, respectively, $\D\cap\O_j$}\cr
(b)\ &E_\D(x)\defi\,\max_{q_i\in\D}\,
\big(\frac{{m\dot q}^2}{2} +
\psi(q)\big)/\f_0\cr
(c)\ &U_\D(x)=\frac1{2} \sum_{q_i,q_j\in\D,\, i\ne j}
\f(q_i-q_j)/\f_0\cr
(d)\ &V_\D(x)= \max_{q_i\in\D} \frac{|\dot q_i|}{v_1}
\cr
}\Eq{e2.8}\ee

The symbol ${\BB}(\x,R)$ will denote the ball centered at $\x$ and
with
radius $R \,r_\f$. With the above notations the {\it local
dimensionless energy} in ${\cal B}(\x,R)$ will be defined as
$W(x;\x,R) \defi$ $ E_{\BB(\x,R)}(x)+ $$U_{\BB(\x,R)}(x)+N_{\BB(\x,R)}(x)$
or, more explicitly,

$$
\eqalignno{ &W(x;\x,R) \defi\frac1{\f(0)}\sum_{q_i\in {\cal
B}(\x,R)}\Big(\frac{m\dot q_i^2}2+\ps(q_i)&\eq{e2.9}\cr 
&+\frac12\sum_{q_i,q_j\in\BB(\x,R),\, i\ne j}
\f(q_i-q_j)+\f(0) \Big)\cr}$$
Let $\log_+ z\defi\max\{1,\log_2|z|\}$, $g_\z(z)=(\log_+ z)^\z$ and

\be
\EE_\z(x)\defi \sup_{\x}\sup_{R> g_\z(\frac{\x}{r_\f})} \frac{W(x;\x,R) }{R^d}
\Eq{e2.10}\ee

If $\HH$ is the space of the locally finite configurations ({\it i.e.}
containing finitely many particles in any finite region) and let
${\cal H}_\z\subset\HH$ be the configurations with

\be\eqalign{
(1) \ &\ {\cal  E}_\z(x)<\infty,\qquad
(2) \ \frac{K_{j,\L}}{|\L\cap\O_j|}\,
>\frac12\frac{\d_j\,d}{2\b_j}\cr}\Eq{e2.11}\ee
for all $\L={\cal B}(O,L)$ large enough and for $\d_j,T_j$, given by
Eq.\equ{e2.2}. Let $N_{j,\L}$, $U_{j,\L}, K_{j,\L}$ denote the number
of particles and their potential or kinetic energy in
$\O_j\cap\L$. Each set ${\cal H}_\z$ has $\m_0$-probability $1$ for
$\z\ge1/d$, see Appendix A,B.

\def\SEC{Equivalence}
\section{Equivalence: isoenergetic versus
Hamiltonian}\label{sec3}
\iniz

It can be expected (and proved here if $d=1,2$) adapting to the
present situation a conjecture proposed in \cite{Ga008d}, that the
following property holds for the time evolutions
$x^{(\L,a)}_i(t)\defi(q_i^{(\L,a)}(t),\dot q_i^{(\L,a)}(t))$, $a=0,1$,
of an initial configuration $x$:

\vskip3mm
\noindent {\bf Local dynamics} {\it Let $d=1,2,3$. Given $\Th>0$, with
  $\m_0$--probability $1$ then for $t\in [0,\Th]$,
\\ (1) The limits
$x^{(a)}(t)\defi $ $\mathop{\lim}\limits_{\L\to\infty}$ $ x^{(\L,a)}(t)$
(``thermo\-dyna\-mic limits'') exist for all $t\le \Th$ and $a=0,1$.
\\ (2) For $t\le\Th$, $x^{(\L,1)}(t)$ satisfies the second of
Eq.\equ{e2.11}.
\\
(3) The function $t\to x^{(0)}(t)$ solves uniquely
the Hamiltonian equations in a subspace of $\HH$ to which also
$x^{(1)}(t)$ belongs (explicit, sufficient, bounds are described in
theorem 7).} \vskip3mm

\noindent{\it Remarks:} (a) The limits of $x^{(\L,a)}(t)$, as
$\L\to\infty$, are understood in the sense that for any ball $\D$
whose boundary does not contain a particle of $x^{(0)}(t)$ the labels
of the particles of $x^{(a)}(t)$ and those of the particles in
$x^{(\L,a)}(t)$ which are in $\D$ are the same and for each $i$ the
limits $\lim_{\L\to\infty} (q_i^{(\L,a)}(t), \dot q_i^{(\L,a)}(t))$
exist and are continuous, together with their first two derivatives
for each $i$.
\vskip2mm

\noindent(b) Uniquess in (3) can be given several meanings. The
simplest is to require uniqueness in the spaces $\HH_\z$ for
$\z\ge1/d$ fixed: and theorem 9 shows that for $d=1,2$ one could
suppose such simpler property. However our result is more general and
we have left deliberately undetermined which subspace is meant in (3)
so that the determination of the subspace has to be considered part of
the problem of establishing a local dynamics property. The
generality might become relevant in studying the case $d=3$, where
even in equlibrium there is no proof that the evolution of data in
$\EE_\z$ remains in the same space.  \vskip2mm

\noindent(c) Recalling the characteristic velocity scale (namely
$v_1=\sqrt{{2\f(0)}/{m}}$), the initial speed of a particle located in
$q\in\RRR^d, \,|q|>r_\f$, is bounded by $v_1\sqrt{\EE}
g_{1/d}(q/r_\f)^{\frac{d}2}$; and the distance to the walls
of the particle located at $q$ is bounded by $(\sqrt{\EE}
g_{1/d}(q/r_\f)^d)^{-1/\a}\,r_\f $. \vskip2mm

\noindent(d) Hence for $|q|$ large they are, respectively, bounded
proportionally to $[(\log |q|/r_\f)^\frac1d]^{\frac{d}2}$ and $[(\log
|q|/r_\f)^\frac1d]^{\frac1\a}$: this says that locally the particles
have, initially, a finite density and reasonable energies and velocity
distributions (if measured on boxes of a ``logarithmic
scale''). The theorem 9 in appendix B will show that this property
remains true for all times, with $\m_0$ probability $1$.\vskip2mm

\noindent(e) An implication is that Eq.\equ{e2.6} has a meaning at
time $t=0$ with $\m_0$--probability $1$ on the choice of the initial
data $x$, because $\EE(x)<\infty$.  \vskip2mm

\noindent(f) The further property that the thermostats are {\it efficient}:
{\it i.e.} the work performed by the external non conservative forces
is actually absorbed by the thermostats in the form of heat $Q_j$, so
that the system can eventually reach a stationary state, will not be
needed because in a finite time the external forces can only perform a
finite work (if the dynamics is local).
\vskip2mm

\noindent (g) It has also to be expected that, with
$\m_0$--probability $1$, the limits in item (2) of Eq.\equ{e2.11}
should exist and be equal to $\frac{d\,\d_j}{2\b_j}$ for almost
all $t\ge0$ respectively: this is a question left open (as it is not needed
for our purposes).  
\vskip2mm

{\it Assuming the local dynamics} property, equivalence, {\it i.e.}
the property $x^{(0)}(t)\equiv x^{(1)}(t)$ for any finite $t$, can be
established as in \cite{Ga008d}. This is recalled in the next few
lines of this section.

In the thermostatted case, with $\L$--regularized motion, it is
\be |\a_j(x)|=\frac{|\dot{\V X}_j\cdot\partial_j U_{0,j}(\V
X_0,\V X_j)|}{
\dot {\V X}_j^2}\Eq{e3.1} \ee
The force between pairs located in $\O_0,\O_j$ is bounded by
$F{\,{\buildrel def\over=}\,} \max|\partial \f(q)|$; the numerator of
Eq.\equ{e3.1} can then be bounded by $F N_0 \sqrt{\lis
N_\Th}\sqrt{2K_j/m}$ where $N_0$ is the number of particles in $\CC_0$
and $\lis N_\Th$ bounds the number of thermostat particles that can be
inside the shell of radii $D_0$, $D_0+r_\f$ for $0\le t\le \Th$
(having applied Schwartz' inequality).

Remark that the bound on $\lis N_\Th$ exists by the local evolution
hypothesis (see (1) and remark (a)) but, of course, is not uniform
in the initial data $x$).

For $0\le t\le\Th$ and for large enough $\L$, by Schwartz' inequality,

\be |\a_j|\le \frac{\sqrt{m}F N_0\sqrt{\lis N_\Th}}{\sqrt{2
    K_{j,\L}(x^{(1, \L)}(t))}}\le\frac{\sqrt{m}F N_0\sqrt{\lis
    N_\Th}}{\sqrt{|\O_j\cap \L| \d_j d/2\b_j}}\Eq{e3.2} \ee
having used property (2) of the local dynamics assumption; letting
$\L\to\infty$ it follows that $\a_j\tende{\L\to\infty}0$.

Taking the limit $\L\to\infty$ of Eq.\equ{e2.6} {\it at fixed $i$},
this means that, {\it with $\m_0$--probability $1$}, the limit motion as
$\L\to\infty$ (with $\b_j, \l_j,\,j>0, $ constant) satisfies

\be
q_i(t)=q_i+\int_0^t \dot q_i(t')dt',\, \dot q_i(t)=\dot q_i+\int_0^t
F_i(t'') dt''\Eq{e3.3}\ee
{\it i.e.} Hamilton's equations, see Eq.\equ{e2.6}; and the solution
to such equations is unique with probability $1$, (having again used
assumption (3) of the local dynamics). In conclusion\vskip2mm

\0{\bf Theorem 1: }{\it If the dynamics is local in the above sense
then in the thermodynamic limit the thermostatted evolution, within
any prefixed time interval $[0,\Th]$, becomes the Hamiltonian
evolution at least on a set of configurations which have probability
$1$ with respect to the initial distribution $\m_0$, in spite of the
non stationarity of the latter.}  
\vskip2mm

Suppose, in other words, that the initial data are sampled with the
Gibbs distributions for the thermostat particles (with given
temperatures and densities) and with an {\it arbitrary distribution}
for the finite system in $\O_0$ with density with respect to the
Liouville volume 
(for instance with a Gibbs distribution at temperature $T_0$ and
chemical potential $\l_0$ as in Eq.\equ{e1.6}).

Then, {\it in the thermodynamic limit}, the time evolution is the same
that would be obtained, in the same limit $\L\to\infty$, via a
isoenergetic thermostat acting in each container $\O_j\cap\L$ and keeping
its total energy (in the sector with $N_j$ particles) constant and with
a density equal (asymptotically as $\L\to\infty$) to $e_j$.

The difficulty of proving the locality property (2) cannot be
underestimated, although it might seem, at first sight, ``physically
obvious'': the danger is that evolution implies that the thermostat
particles {\it grind to a stop} in a finite time converting the
kinetic energy entirely into potential energy. The consequence would
be that $\a_j$ becomes infinite and the equations even ill defined.
As a consequence it is natural to expect, as stated in the local
dynamics assumption, only a result in $\m_0$--probability. This can be
better appreciated if the following {\it counterexample}, in the
Hamiltonian case, is kept in mind.

Consider an initial configuration in which particles are on a square
lattice (adapted to the geometry): regard the lattice as a set of
adjacent tiles {\it with no common points}. Imagine that the particles
at the four corners of each tile have velocities of equal magnitude
pointing at the center of the tile. Suppose that the tiles sides are
$> r_\f$. If $\f(0)$ is large enough all particles come to a stop in
the same finite time and at that moment all kinetic energy has been
converted into potential energy: at time $0$ all energy is kinetic and
later all of it is potential. Certainly this example, which concerns a
single event that has, therefore, $0$ probability in $\m_0$, shows
that some refined analysis is necessary: the thermostatted evolution
$x^{(\L,1)}(t)$ might be not even be defined because the denominator
in the definition of $\a_j$ might become $0$.  \vskip2mm

It should be stressed that the thermostats models considered here
preserve even at finite $\L$ an important symmetry of nature: {\it
time reversal}: this ceertainly explains the favor that they have
received in recent years in the simulations.

Finally a corollary will be that the non dissipative Hamiltonian
motion and the dissipative thermostatted motions, although different
at finite volume become identical in the thermodynamic limit: neither
conserves phase space volume (measured with $\m_0$) but in both cases
the entropy production rate coincides with the phase space
$\m_0$--volume contraction. 


\def\SEC{Free thermostats}
\section{Free thermostats}\label{sec4}
\iniz

The need for interaction between the particles in order to have a
physically sound thermostat model has been stressed in
\cite{Ru99,Ru01} and provides a measure of the importance of the
problems met above.

The above discussion is heuristic because the local dynamics
assumption is not proved. However if the model is modified by keeping
only the interaction $\f$ between the test particles and between test
particles and thermostat particles, but suppressing the interactions
between particles in the same $\O_j,\,j>0$, and, furthermore,
replacing the wall potentials by an elastic collision rule. {\it I.e.}
supposing $U_j(\V X_j)\equiv 0,\, j>0$, together with the collision
rule, the analysis can be further pursued and completed. This will be
referred as the ``{\it free thermostats}'' model.

It can be remarked that in the Hamiltonian case this is the classical
version of the Hamiltonian thermostat models that could be completely
treated in quantum mechanics, \cite{FV63}.

Let $\L_n$ be the ball ${\cal B}(O,2^n)$ of radius $2^n r_\f$ and
$n\ge n_0$ be such that $2^{n_0}r_\f\ge D_0+r_\f$; if $\lis N$ bounds
the number of particles in the ball $D_0+r_\f$ up to an arbitrarily
prefixed time $\Th$, the first inequality Eq.\equ{e3.2} and the
supposed isoenergetic evolution (which in this case is {\it also}
isokinetic)

\be |\a_j|\le N_0 F \sqrt {\frac {\lis N}{2K_j/m}}\le
\frac{N_0F}{\sqrt{d\,k_B T_j/m}}\defi\ell.\Eq{e4.1}\ee
It follows that, for $\z\ge1$, the speed of the particles initially in
the shell $\L_{n}/\L_{n-1}$ with radii $ 2^nr_\f,2^{n+1}r_\f$ will
remain within the initial speed by, at most, a factor
$\l^{\pm1}=e^{\pm \ell \Th}$.  The initial speed of the latter
particles is bounded by, see Eq.\equ{e2.10},

\kern-3mm
\be V_n= v_1 \sqrt{\EE_\z(x)}\,n^{\frac12 \z d} \Eq{e4.2}\ee
Hence if $n(\Th)$ is the smallest value of $n$ for which the
inequality $2^nr_\f-V_n\,\l\,\Th<D_0+r_\f$ {\it does not hold} no
particle at distance $> 2^{n(\Th)+1}r_\f$ can interact with the test
system.

This means that $\lis N\le \EE_\z(x)\, 2^{\,(n(\Th)+1)\,d}$ and the
dynamics $x^{(n,a)}(t)$ becomes a finitely many particles dynamics
involving $\le N_0+\lis N$ particles at most.

From the equations of motion for the $N_0+\lis N$ particles we see
that their speed will never exceed

\be V_\Th=(\lis V +F\,N_0\,\lis N\,\Th) \,\l\Eq{e4.3}\ee
if $\lis V$ is the maximum of their initial speeds. In turn this means
that for $n$ large enough a better bound holds on $\a_j$,

\be|\a_j \dot q_i|\le \frac{N_0 \lis N V_\Th^2 F}{\o_{j,n} 2^{dn}r_\f^d\d k_B
  T/m}\tende{n\to\infty}0\Eq{e4.4}\ee
with $T=\min_{j>0}T_j$ and $\d=\min_{j>0}\d_j$ and $\o_{j,n}
2^{dn}r_\f^d$ bounds below (for suitable $\o_{j,n})$ the volume of
$\O_j\cap\L_n$.

Hence, for $a=0,1$, it is $\lim_{n\to\infty}
x^{(n,a)}(t)=x^{(0)}(t)$, and {\it also} the dynamics is local in the
above sense. This completes the analysis of free
thermostats and proves:

\vskip2mm
\0{\bf Theorem 2:} {\it Free isoenergetic and Hamiltonian thermostats
  are equivalent in the thermodynanic limit}
\vskip2mm

Notice that essential use has been made of the property
that, in absence of interaction among pairs of thermostat particles
and in presence of perfectly elastic walls, isokinetic and
isoenergetic dynamics coincide: so the denominators in Eq.\equ{e4.1}
are constant.

It would be possible to consider non rigid walls, modeled by a soft
potential $\ps$ diverging near them. We do not perform the analysis
because it is a trivial consequence of the analysis that follows. We
have chosen the example of this section because it pedagogically
illustrates well the simplest among the ideas of the coming analysis.

\def\SEC{Kinematics}
\section{Kinematics}\label{sec5}
\iniz

The proof of the local dynamics property will require controlling the
maximal particles speeds, the number of particles interacting with any
given one as well as their number in any finite region. This will be
achieved by proving bounds on the local energies $W(x;\x,R)$,
Eq.\equ{e2.9}.

In this section we shall prove bounds at time $0$, see Eq.\equ{e5.3}
below as a preparation to the next section where we shall use energy
conservation to extend the bounds to positive time.

To study general thermostats in dimension $d$ consider the
dimensionless sum of the energy, measured in units of $\f_0$, and the
particle number in the ball ${\cal B}(\x,R)$, with center $\x$ and
radius $r_\f\,R$ as defined in Eq.\equ{e2.9} and denoted
$W(x;\x,R)$.

The potential $\f$ is superstable so that the number $N$ of points in
a region $\D$ can be bounded in terms of the potential energy $U$ in
the same region and of $\f_0=\f(0)$ and $\lis\f>0$ (defined after
Eq.\equ{e1.4}). This is checked below.
\*

In fact, by the dimensionless energy definition in Eq.\equ{e2.9},
$W\ge (\frac {U}{\f_0}+N)\ge \frac{\lis \f}{2\f_0} \sum _p N_p^2$ with
the sum running over labels $p$ of disjoint boxes of diameter
$\frac{r_\f}2$ covering $\D$ and containing $N_p\ge0$ particles (in
particular: $N=\sum_p N_p$), hence over $\ell
\le{|\D|}{(2\sqrt{d}/r_\f)^{d}}$ terms.  By the Schwartz' inequality
$\sqrt{{\lis \f}/{2\f_0}} N\le \sqrt{W\,\ell}$ gives a bound of the
total number of particles in a region $\D$ in terms of the local
dimensionless energy $W$:

\be N_\D\le C \frac{\sqrt W}{\sqrt{|\D|}},\qquad C\defi
\big(\frac{2\f_0}{\lis\f}\big)^{\frac12}\Eq{e5.1}\ee
This is the well known ``superstability estimate'' (derived in our
simplifying assumptions of $\f\ge0$ and finite range).
\*

Calling $\EE\defi\EE_{1/d}(x)$, Eq.\equ{e2.10}, consider a
sequence of balls $\L_n={\cal B}(O,2^n)$, of radii $L_n= 2^n r_\f$
with $n\ge n_0$: so that $2^{n_0} r_\f> D_0+r_\f$ and all $\L_n$
enclose the test system and the particles interacting with it.  Given
a configuration $x$ define $N(x;\x,R)$ the number of particles in the
ball of radius $R\,r_\f$ centered at $\x$ and, given $x$,

\be\eqalign{
(1)\,\, &  V_n = \,\hbox{ the maximum velocity in}\ \L_n\cr
(2)\,\, &  \r_n= \,\hbox{ the minimum distance to }\
\partial (O_j\cap\L_n)\cr
(3)\,\, &  \NN_n=\, \max_{q_i\in\L_n} N(x;q_i,1)\cr}\Eq{e5.2}\ee
After the definition of
$W,\EE$, the {\it initial speed} of a particle in the ball $\L_n,
\,n\ge1$, and its distance to the walls will be bounded above and
respectively below by $v_n,\r_n$ with, see definitions Eq.\equ{e2.7},

\be V_n=v_1\, (n\EE)^{\frac12},\quad \r_n = \frac{r_\ps}{(n\,
{\EE})^{1/\a}},\quad \NN_n\le C(n\EE)^{\frac12} \Eq{e5.3}\ee
under the assumption that the wall potential has range $r_\ps$ and is
given by Eq.\equ{e1.2}; the last inequality is a consequence of the
definition of $W$ and of the above mentioned superstability.
\*

\0{\it Constant convention:} From now on we shall encounter various constants
that are all computable in terms of the data of the problem (geometry,
mass, potentials, densities, temperatures and the (arbitrarily)
prefixed time $\Th$) as in the above Eq.\equ{e5.1} which gives a
simple example of a computation of a constant. To avoid proliferation
of labels all constants will be positive and denoted $C,C',C'',\ldots,
B,B',\ldots$ or $c,c',c'', \ldots, b,b',b'',\ldots$: they have to be
regarded as functions of the order of appearance, non decreasing the
ones denoted by capital letters and non increasing the ones with lower
case letters; furthermore the constants $C,\ldots,c,\ldots$ may also
depend on the parameters that we shall name $\EE$ or, in
Sec.\ref{sec9}, $E$ and will be again monotonic non decreasing or non
increasing, respectively, as functions of the order of appearance and
of $\EE$ or $E$.
\*

Consider motions, evolving for times $0\le t\le \Th$, or in the
thermostatted case for $0\le t\le \min\{t_{\L_n}(x),\Th\}$, from an
  initial configuration $x$ following the {\it $\L_n$-regularized}
  evolution of Sec.2 with $n$ fixed (see comment (6), p.\pageref{t}). Define

\be R_n(t)\defi n^{\frac{1}d} +\int_0^t V_n(s) \frac{ds}{r_\f},
\Eq{e5.4}\ee
where $R_n(0)=g_{1/d}(2^n)=n^{1/d}$ and $V_n(s)$ is the
maximum speed that a moving particle can acquire in the time interval
$[0,s]$ under the $\L_n$-regularized evolution.

The choice of $R_n(0)=n^{\frac1{d}}$ is made so that it will be
possible to claim that $W(x(0);O,R(0))\le \EE(x(0)) n$ with
$\m_0$--probability $1$, see Eq.\equ{e2.9},\equ{e2.10} and appendix A.

The dimensionless quantity $R_n(t)$ will also provide a convenient
upper bound to the maximal distance a moving particle can travel
during time $t$, in units of $r_\f$, following the $\L_n$--regularized
motion.

It will be necessary to estimate the total energy and number of
particles in a ball of radius $R_n(t)r_\f$ around $\x$ assuming the
particles to move with the $\L_n$--regularized equations.
Consider the ball around $\x\in\RRR^d$ of radius

\be R_n(t,s)\defi R_n(t)+\int_s^t \frac{V_n(s)}{r_\f}\,ds\ge
1.\Eq{e5.5}\ee
This is a ball whose radius shrinks as $s$ increases between $0$ and
$t$ at speed $V_n(s)$: therefore no particle can enter it. Abridging
$x^{(\L_n,a)}(\t)$ by $x(\t)$, this can be
used to obtain a bound on the size of $W(x(\t);\x,R_n(t))$ in
terms of the initial data $x(0)=x$ and of

\be W_n(x,R)\defi\sup_{\x} W(x_n;\x,R).\Eq{e5.6}\ee
if $x_n$ denotes the particles of $x$ in
$\wt\L_n\defi\BB(O,2^n+1)$, {\it i.e} at distance $\le r_\f$ from $\L_n$.
\vskip2mm

{\it The analysis in the following Sec.\ref{sec6},\ref{sec7} is
taken, with a minor adaptation effort, from the version in {\rm
\cite[p.34]{CMP000}} of an idea in {\rm \cite[p.72]{FD977}} and is
repeated here only for completeness.}  \vskip2mm

Let $\ch_\x(q, R)$ be a smooth function of $q-\x$ that has value $1$
in the ball $\BB(\x,R)$ and decreases radially to reach $0$ outside
the ball $\BB(\x,2R)$ with gradient bounded by $(r_\f\,R)^{-1}$. Let
also

\be\eqalign{&\widetilde W_n(x;\x,R)\defi\frac1{\f_0}
\sum_{q\in\wt\L_n}\ch_\x(q,R)\cr &\cdot \big(\frac{m\dot q^2}2+\ps(q)+
\frac12\sum_{q'\in\wt\L_n}\f(q-q')+\f_0\big).\cr}\Eq{e5.7}\ee
Denoting $B$ an estimate of how many balls of radius $1$ are needed to
cover a ball of radius $3$ (a multiple of the radius large enough for
later use in Eq.\equ{e6.3}) in $\RRR^d$ so that every pair of points
at distance $<1$ is inside at least one of the covering balls, it
follows that $W(x;\x,2R)\le B\, W(x,R)$, see Eq.\equ{e5.5}, so that
for $\x\in\L_n$:

\be \eqalign{& W(x_n;\x,R)\le\widetilde W_n(x;\x,R) \le
W(x_n;\x,2R),\cr
& \widetilde W_n(x;\x,R)\le \,B\, W_n(x,R)\cr}\Eq{e5.8}\ee
Although $W$ has a direct physical interpretation $\widetilde W$
turns out to be mathematically more convenient and, for our purposes,
equivalent by Eq.\equ{e5.8}.

\def\SEC{Energy bound}
\section{Energy bound}\label{sec6}
\iniz

\subsection{Hamiltonian systems}
Considering $\widetilde W(x(s);\x,R_n(t,s))$, for $0\le s\le t\le \Th$,
it follows that

$$\eqalignno{ &\frac{d}{ds} {\widetilde W}_n(x(s);\x,R_n(t,s))\le
\frac1{\f_0}\sum_{q\in\wt\L_n} \ch_\x(q(s),R_n(t,s))\cr
&\cdot\frac{d}{ds} \big(\frac{m\dot q(s)^2}2+\ps(q(s))+
\frac12\sum_{q'\in\wt\L_n}\f(q(s)-q'(s))\big)&\eq{e6.1}\cr}$$
because the $s$--derivative of $\ch_\x(q(s),R_n(t,s))$ is $\le0$ since
no particle can enter the shrinking ball $\BB(\x,R(r,s))$ as $s$
grows: {\it i.e.} $\ch_\x(q(s),R_n(t,s))$ cannot increase.

In the Hamiltonian case a computation of the derivative in
Eq.\equ{e6.1} leads, with the help of the equations of motion and
setting $\ch_\x(q(s),R_n(t,s))\equiv\ch_{\x,q,t,s}$, to

\be\eqalign{ & \frac{d}{ds} \widetilde W_n(x(s);\x,R_n(t,s))\le
\sum_{q\in\O_0} \dot q(s) \F(q(s))\ch_{\x,q,t,s}\cr &
-\sum_{q\in
\L_n\atop q'\in\wt \L_n}\big(\ch_{\x,q,t,s}\,- \ch_{\x,q',t,s}\,\big)
\frac{\dot q(s)\partial_q \f(q(s)-q'(s))}{2\f_0}\cr
}\Eq{e6.2}\ee
where the dot indicates a $s$--derivative and it has been kept in mind
that positions and velocities of the particles outside $\L_n$ are
considered, in the $\L_n$--regularized dynamics, to be time
independent.

Since the non zero terms have $|q(s)-q'(s)|<r_\f$ and the derivatives
of $\ch$ are $\le (r_\f R_n(t,s))^{-1}$ and $|\dot q|,|\dot q'|\le
V_n(s)=r_\f |\dot R_n(t,s)|$ it follows, setting $F=\max(|\partial
\f|+|\F|)$,

\be \eqalign{ &\frac{d}{ds} \widetilde W_n(x(s);\x,R_n(t,s))\le \frac{F
v_1}{\f_0} \widetilde W_n(x(s);\x,\frac{D_0}{r_\f}) \cr & +\frac{F
r_\f}{\f_0} \frac {|\dot R_n(t,s)|}{R_n(t,s)}B\, \widetilde W_n(x(s);\x,2
R_n(t,s)+1)\cr &\le B^2\, \frac{F v_1}{\f_0} (\frac{r_\f}{
v_1}\frac {|\dot R_n(t,s)|}{R_n(t,s)}+1)\widetilde
W_n(x(s);R_n(t,s))\cr}\Eq{e6.3}\ee
where $\widetilde W_n(x;R)$ is defined in analogy with Eq.\equ{e5.6}.

\0Eq.\equ{e6.3}, $R_n(t,s)/R(t,0)\le 2$ and $W_n(x(s),\x, R_n(t,s))$
$\le W_n(x(s),R_n(t,s))$ imply the inequality

\be W_n(x(s),R_n(t,s))\le e^\G\,
W_n(x(0), R_n(t,0)), \Eq{e6.4}\ee
with $\G$ $\defi{(\frac{Fr_\f}{\f_0}\log2+\frac{3F
v_1}{\f_0})\,\Th\, B}$.

\subsection{Thermostatted systems}

In the thermostatted dynamics case, Eq.\equ{e6.2},\equ{e6.3}
have to be modified by adding to the r.h.s the work, per unit time,
done by the thermostatic forces (measured in units of $\f_0$) which
is, see Eq.\equ{e1.5},

\be \frac{\sum_i^* \dot q_i(s)F_i(x(s))}{\f_0\sum_i \dot
  q_i(s)^2}\sum_{q_i\in \BB(\x,R_{n}(t,s))} \kern-2mm
  \ch_\x(q_i,R_{n}(t,s))\, \dot q_i(s)^2 \Eq{e6.5}\ee
here the $*$ means restriction of the sum to the particles $q_i(t)$
  in $\O_j$ and within distance $r_\f$ from the boundary of $\O_0$,
  which is a ball of radius $D_0$. Then Eq.\equ{e6.5} is bounded by

\be N_0\, \frac{F v_1}{\f_0}\, \widetilde
  W_n(x(s),R_n(t,s))\Eq{e6.6}\ee
because the bounds of the various
terms in Eq.\equ{e6.5} can be obtained as:
\vskip2mm

\noindent(a) the $\sum_i \dot q_i(s) F_i(q(s))$ by (Schwarz'
inequality) $\le N_0 F (\sum^*_i \dot q_i^2)^{\frac12} (\sum^*_i
1)^{\frac12}$ with $F=\max|\partial \f|$. Leaving  aside the factor
$FN_0$ the rest is bounded above proportionally to
$\widetilde W_n(x(s),R_n(t,s))$.  \\
(b) the kinetic energy in the last sum in Eq.\equ{e6.5} is also
bounded by the total kinetic energy and is compensated by the
denominator.  \\
(c) $N_0$ is bounded in terms of $\EE$ (and constant in time).
\vskip2mm

Therefore, given $n$, the new inequality which replaces Eq.\equ{e6.3}
in the thermostatted case gives a bound of the $s$-derivative $ \dot
{\widetilde W}_n\defi$ $ \frac{d}{ds} {\widetilde
W}_n(x(s);\x,R_{n}(t,s))$ in terms of, see Eq.\equ{e5.6},\equ{e5.8},
${\widetilde W}_n$ $\defi$ $ \sup_\x{\widetilde W}_n(x(s);\x,R_{n}(t,s))$,
namely

\be \dot {\widetilde W}_n\le C\, \big(\frac{\dot
R_{n}(t,s)}{R_{n}(t,s)}\,+(N_0+1)\, \frac{\lis F v_1}{\f_0}
\,\Big)\,\widetilde W_n\,,\Eq{e6.7}\ee

The second addend in Eq.\equ{e6.7} is bounded, by (c) above. The
differential inequality Eq.\equ{e6.7} (together with
Eq.\equ{e5.6},\equ{e5.8}) implies, for suitable $C',C^2$ functions of
$\EE$:

\be W_n(x(t),R_n(t))\le C' W_n(x(0),R_n(t)) \le C^2
R_{n}(t)^d\Eq{e6.8}\ee
for $t\le \min\{t_{\L_n}(x),\Th\}$ (see p.\pageref{t}, item
(6)). Hence the bound on the speed, for instance for $d=2$,
$\frac{V_n(t)}{v_1}\le C R_n^{{d}/2}$ with $R_n=R_n(\Th)$.

\subsection{The bound}

Therefore the following {\it energy bound} holds:
\vskip2mm

\0{\bf Theorem 3:} {\it For a suitable constant $C$ the
following energy bound holds for the $\L_n$--regularized Hamiltonian
or thermostatted dynamics and for $t\le\Th$ or
$t\le\min\{t_{\L_n}(x),\Th\}$, respectively,

\be W_n(x^{(n,0)}(t), R_n(t))\le \,C^2\, R_n(t)^d \Eq{e6.9}\ee
 and $n$ large enough and $C>0$ (depending only on $\EE$).
} \vskip2mm

\noindent{\it Remark:} the inequality holds for all $d$'s and in the
Hamiltonian case the constant $C$ can be taken $\EE$--independent, see
Eq.\equ{e6.4}.
\vskip2mm

By the definition Eq.\equ{e2.9} of $W$ it follows that

\be V_n(s)\le v_1\,C\, R_n(s)^{\frac d2}\Eq{e6.10}\ee
and going back to Eq.\equ{e5.5}, and solving it,
$R_n(t)$ is bounded proportionally to $R_n(0)=n^{1/d}$.

Calling $\r_n(t),V_n(t),\NN_n(t)$ the quantities in Eq.\equ{e5.2} for
$S^{(n,1)}_tx$ the following bounds can be formulated simultaneously
for the thermostatted and Hamiltonian cases and extend to positive
times the estimates in Eq.\equ{e5.3}.

\vskip2mm \0{\bf Theorem 4:} {\it If $d\le 2$ the maximal velocity
  $V_n(t)$ and the maximal displacement $R_n(t) \,r_\f$, in the
  $\L_n$--regularized motion up to time $t\le\min\{\Th,t_{\L_n}(x)\}$ in
  the thermostatted case or $t\le \Th$ in the Hamiltonian case satisfy:

\be \eqalign{
R_n(t)\le& \,C\, n^{\frac{1}d},\quad \qquad
V_n(t)\le v_1 C\, n^{{1}/2}
\cr
\r_n(t)\le& \, C^{-1} n^{-\frac1\a},\quad
\NN_n(t)\le C n^{\frac12} \cr}
\Eq{e6.11}\ee
with $C>0$ (a suitable function of $\EE$).}
\vskip2mm

The use of Eq.\equ{e6.10} to solve Eq.\equ{e5.4}, which then implies
Eq.\equ{e6.11}, will force the restriction on the dimension to be
$d\le2$. {\it However if the Eq.\equ{e6.11} are assumed also for $d=3$
the following analysis and results would hold true} also for $d=3$,
except a ``minor'' modification of the extra result in theorem 9 in
appendix C; therefore we keep in the following the dimension $d$
generic.

\def\SEC{Entropy bound and phase space contraction}
\section{Entropy bound and phase space contraction}\label{sec7}
\iniz

The kinetic energy density $K_{j,n}(x)/|\O_j\cap\L_n|$, {\it i.e.}
the kinetic energy contained in the $j$-th thermostat at time $t$ in
the $\L_n$--regularized thermostatted or Hamiltonian dynamics divided
by its volume, will initially have a value as close as wished to $
\d_j \frac{d}2 k_B T_j$ for $n$ large enough. So that
$\frac{K_{j,n}}{\f_0}{\buildrel def\over \simeq} \,\k_j\, 2^{dn}$ if $n$
is large enough: because this is an event which has probability $1$
with respect to $\m_0$.

Therefore if the infinite volume $\m_0$--average value of
$\frac{K_{j,n}(x)}{\f_0}$ is denoted $\k_j
2^{n\,d}$ then there is $n(x)$ such that

\be \frac{K_{j,n}(x)}{\f_0}> \frac12 \k\,2^{n\,d},\qquad
\forall\ n\,\ge\, n(x).\Eq{e7.1}\ee
where $\k=\min_{j>0} \k_j$. The notation $x(t),x(s),\ldots$ will be
temporarily used below for simplicity instead of
$x^{(n,a)}(t),x^{(n,a)}(s),\ldots$.  The equations of motion are now
Eq.\equ{e1.3}, or \equ{e2.6}, with $a=1$.

With $\m_0$--probability $1$ it is $\EE(x),n(x)<\infty$.  Let
$\X_{E,h}$ be the set of configurations for which $\EE(x)\le
E,n(x)=h$. Let $\n_h(x)$ be the smallest $n\ge h$ such that the
event $\min_{j>0}\frac{K_{j,n}(x(t))}{\f_0}= \frac12 \k\,2^{n\,d}$ is realized
for $t=t_{n,h}(x)<\Th$ but not earlier.
\*

\0{\it Remark:} the definition of $\X_{E,h}$ implies that the time
$t_{\L_n}(x)$ (see comment (6), p.\pageref{t}) is, for $x\in\X_{E,h}$,
certainly $>t_{n,h}(x)$. \*

Let $\X_{E,h,n}$ be the set of the $x\in\X_{E,h}, \,
\n_h(x)=n$.  It will be shown that $\sum_{n\ge h}
\m_0(\X_{E,h,n})<\infty$: hence with $\m_0$--probability $1$ a point $x$
will be out of $\X_{E,h,n}$ for all $n$ large enough and Eq.\equ{e7.1}
will hold for all $t\le \Th$ with $\m_0$--probability $1$.

For $x\in \X_{E,h,n}$, consider $Q_j$ as in Eq.\equ{e1.5}.  {\it
  Notice that $Q_j$ has the physical interpretation of the heat ceded
  per unit time to the $j$-th thermostat by the system.}  \vskip2mm

If $N_0$ is the number of particles in $\O_0$, $\NN_n(t)$, $V_n(t)$
are as in Eq.\equ{e6.11}, $F$ is the maximum of $|\partial \f|$, and
$\lis N$ is the maximum number of particles within $r_\f$ of the
boundary of $\O_0$, then $|Q_j|\le N_0\lis N F V_n(t)\le C
n^{\frac12+\frac12}$ because $V_n(t)$ is bounded by Eq.\equ{e6.11}
proportionally to $n^{1/2}$, $\lis N$ is bounded
proportionally to the (finite) number of balls of radius $r_\f$ needed
to cover the ball of radius $D_0+r_\f$ times the $\NN_n(t)$ in
Eq.\equ{e6.11} ({\it i.e} also proportionally to $n^{\frac12}$): the
estimates hold for $t\le \Th$ in the Hamiltonian case and for
$t\le t_{h,n}(x)\le \Th$ in the thermostatted case..

The phase space contraction $\s=\s(x)$, see Eq.\equ{e2.3}, is

\be \s=\sum_{j>0}d N_j \frac{Q_j}{2K_{j,n}(x,t)} (1-\frac1{d N_j})
+\b_0 Q_0
\Eq{e7.2}\ee
in the thermostatted case, if $Q_0\defi -\sum_{q\in \O_0}$
$\sum_{j>0}$ $ \dot {\bf X}_0\cdot$ $\partial_{\bf X_0} U_{0,j}(\V
X_0,\V X_j)+\dot{\V X}_0\cdot\BF$;
in the Hamiltonian case,  it is:

\be \s(x)=\sum_{j\ge0}\b_j Q_j\Eq{e7.3}\ee
It has to be kept in mind that there is contraction of phase space
even in the fully Hamiltonian case when Liouville's theorem holds
(already for the regularized dynamics): this is no contradiction
because the phase space volume is measured with the volume defined by
$\m_0$.

Since $Q_0$ can be estimated in the same way as $Q_j$, above, it
follows that the integral $|\int_0^{t_{h,n}(x)}\s(x^{(n,a)}(t))|$ is,
{\it in either cases}, also uniformly bounded (in $\X_{E,h}$) by

\be \lis \s\Th n,\qquad \forall \, n\ge h \Eq{e7.4}\ee
where $\lis\s$ is a suitable non decreasing function of $E$. This is
a first {\it entropy estimate}; it is rather far from optimal and it
will appear, Sec.\ref{sec9}, that it will be essential to impove it.

Therefore a volume element in $\X_{E,h,n}$ contracts at most by
$e^{-\lis\s\Th n}$ on the trajectory of $\m_0$-almost all
points $x\in\X_{E,h,n}$ up to the stopping time $t_{h,n}(x)$.

Effectively this means that the distribution $\m_0$ can be treated as
an invariant one for the purpose of estimating the probability that
the kinetic energy, in $\O_j\cap\L_n$ of the initial data $x\in
\X_{E,h,n}$, in the time $t_{h,n}(x)$ grinds down to $\frac{\k}2 2^{dn}$
({\it i.e.} to half (say) of the value $\k 2^{nd}$ to which it is
intially very close, if $n$ is large). The estimate can be carried out
via the technique introduced by Sinai, \cite{Si974}, which has been
applied in \cite{MPP975, FD977}; for completeness see the following
appendix D.

Let $D=D_{n}$ be the set of the $x\in\X_{E,h,n}$ which satisfy
$K_{j,\L_n}(x)=\frac12 \k 2^{dn}$ for a given $j>0$ while
$K_{j',\L_n}(x)=\frac12 \k 2^{dn}$ for $j'>0,j'\ne j$. 

Recalling the DLR-equations,
\cite{LR969}, and the classical superstability estimate on the
existence of $b>0$ such that $p_n\le e^{-b 2^{dn}}$ bounds the
probability of finding more than $\r 2^{d n}$ particles in
$\L_n\cap\O_j$ if $\r$ is large enough ({e.g.}  $\r> \max_j\d_j$), the
probability $\m_0(\X_{E,h,n})$ can be bounded by $p_n$ (summable in
$n$) plus

$$\eqalignno{ & e^{\lis\s\Th  n}\int \m_0(dq' d\dot q')\sum_{l=1}^{\r 2^{n d}}
\Th \frac{e^{-(\b_j
U_{\L_n,j}(q,q')-\l_j\,l)}}{Z_{\L_n,j}(q')} \frac{dq} {l!}\cr
&\cdot\, e^{-{\b_j P^2}} P^{l\,d-1 } \widehat
P\,\o(l\,d)&\eq{e7.5}\cr}$$
where $q=(q_1,\ldots,q_l)\in (\O_j\cap\L_n)^l$, $P^2=\frac12 \k
2^{dn}$, $U_{\L_n,j}(q,q')$ is the sum of $\f(q-q')$ over the pairs of
points $q_i,q'_\ell\in \O_j\cap\L_n$ plus the sum over the pairs with
$q_i\in\L_n\cap \O_j, q'_\ell\not\in\L_n\cap \O_j $, and
\vskip2mm

\noindent(1) $Z_{\L_n,j}(q')$ is the partition function for the region
$\L_n\cap\O_j$ (defined as in Eq.\equ{e2.1} with the integral over the
$q$'s extended to $\L_n\cap\O_j$ and with the energies
$U_{\L_n}(x,z)$);
\vskip2mm\noindent(2) The volume element $P^{ld-1}dP$ has been changed
to $P^{ld-1}\dot P d\t=P^{ld-2} \widehat P d\t$ where $\widehat P$ is
a {\it short hand} for $\sum_{q,q';\, q\in\L_n} |\partial_q
\f(q-q')|+\sum_{q\in \L_n} |\partial_q\ps(q)|$ so that $P\widehat P$
is a bound on the time derivative of the total kinetic energy $P^2$
contained in $\L_n$ evaluated on the points of $D_n$ (the latter is
$2P\dot P=|\sum_{i,j; q_i\in \L_n} \partial \f(q_i-q_j) (\dot q_i-\dot
q_j)+\sum_{q\in \L_n} \partial_q\ps(q)\dot q|$ hence $\le P^2\widehat
P$).
\vskip2mm\noindent(3) $\o(l\,d)$ is the surface of the unit ball in
$\RRR^{l\,d}$.
\vskip2mm\noindent(4) The factor $e^{\lis\s\Th n}$ takes into account
the entropy estimate, {\it i.e.} the estimate Eq.\equ{e7.4} of the
non-invariance of $\m_0$.   \vskip2mm

The integral over $\t$ in Eq.\equ{e7.5} gives a factor $\Th$ and the
integral can be trivially imagined averaged over an auxiliary
parameter $\e\in[0,\lis\e]$ with $\lis\e>0$ arbitrary (but to be
suitably chosen shortly) on which it does not depend at first. Then if
$P$ is replaced by $(1-\e)P$ in the exponential while $P^{l\,d-1}$ is
replaced by $\frac{((1-\e)P)^{l\,d-1}}{(1-\e)^{\r 2^{nd}-1}}$ the
average over $\e$ becomes an upper bound. Changing $\e$ to $P\e$ ({\it
i.e.} hence $d\e$ to $\frac{dP\e}P=\frac{2dP\e}{\k 2^{dn}}$) the bound
becomes the $\m_0$-average

\be \eqalign{
&\frac{2 e^{\lis\e \r 2^{nd}}}{\lis\e \k 2^{dn}}\langle {\widehat P\,
\ch_{\k,\lis\e}} \rangle_{\m_0} \equiv
\frac{2e^{\lis\e \r 2^{nd}}}{\lis\e \k 2^{dn}}
\langle \widehat P^2\rangle^{\frac12}_{\m_0}\,
\cdot\,\langle \ch_{\k,\lis\e} \rangle^{\frac12}_{\m_0}\cr&\le B\, e^{-b\,
2^{{nd}/2}}\cr} \Eq{e7.6} \ee
where $\ch_{\e,\k}$ is the characteristic function of the set $\big\{
(1-\lis\e)^2\frac\k2 2^{dn}<K_j<\frac\k2 2^{dn}\big\}$. The inequality is
obtained by a bound on the first average, via a superstability
estimate, proportional to $2^{2dn}$ and by the remark that the second
average is over a range in which $K$ shows a large deviation from its
average (by a factor $2$) hence it is bounded above by $e^{-b 2^{nd}}$
with $b$ depending on $\k$ but independent on $\lis\e$ for $n$
large. Therefore fixing $\lis\e$ small enough (as a function of $\k$)
the bound holds with suitable $B,b>0$ and is summable in $n$ (and of
course on $j>0$).

Hence, fixed $h$, with $\m_0$--probability $1$ it is
$K_{j,n}\ge\frac12\k 2^{n\,d}$ (by Borel-Cantelli's theorem) for all
$n$ large enough and $j>0$. As mentioned after Eq.\equ{e7.1} this means that for
all $t\le \Th$ it is $K_{j,n}\ge\frac12\k 2^{n\,d}$ for all $n$ large
enough, with $\m_0$ probability $1$. Therefore the bounds in
Eq.\equ{e6.11} can be assumed, with $\m_0$--probability $1$ also for
the {\it thermostatted and Hamiltonian dynamics} and for all $n$ large
enough.

\vskip2mm
\noindent{\bf Theorem 5:} {\it With $\m_0$--probability $1$ the phase
space contraction $\s(x)$ admits a bound $\lis\s(x)<\lis \s n$ for all
times $t\le \Th$ for the $\L_n$--regularized thermostatted or
Hamiltonian dynamics. Furthermore the kinetic energy
$K_{j,n}(x^{(n,a)}(t))$, $a=0,1$, in the $j$--th thermostat remains
$\ge \k 2^{nd}$ for all $n>n(x)$ for suitable $\k,n(x)>0$. The
constants $\lis\s(x),n(x)$ depend on $x$ only through $\EE(x)$. }
\vskip2mm

This proves item (2) of the local dynamics property.

\def\SEC{Infinite volume Hamiltonian dynamics}
\section{Infinite volume Hamiltonian dynamics}\label{sec8}
\iniz

It remains to check that also the $n\to\infty$ limit of the dynamics
exists in the sense of the local dynamics assumption ({\it i.e.}  the
existence of the limit $x(t)\equiv x^{(0)}(t)\defi\lim_{n\to\infty}
x^{(n,0)}(t)$ and a suitable form of its uniqueness).

The equation of motion, for a particle in the $j$-th container (say),
can be written both in the Hamiltonian and in the thermostatted cases as

\be
\eqalign{& q_i^{(n,a)}(t)=q_i(0)+\int_0^t \big(e^{-\int_0^\t
  a\a_j(x^{(n,a)}(s))ds}\dot q_i(0)\cr
&+
(t-\t)\,e^{-\int_\t^t
  a\a_j(x^{(n,a)}(s))ds}\,f_i(x^{(n,a)}(\t))\big)\,d\t\cr}\Eq{e8.1}\ee
\noindent{}where the label $j$ on the coordinates (indicating the
container) is omitted and $f_i$ is the force acting on the selected
particle divided by its mass (for $j=0$ it includes the stirring
force). The Hamiltonian case is simply obtained setting $a=0$ while
the thermostatted case corresponds to $a=1$.

The existence of the dynamics in the Hamiltonian case, $a=0$, will be
discussed first, proving

\vskip2mm \0{\bf Theorem 6:} {\it If $x\in \HH_{1/d}$ the
thermodynamic limit evolution $x^{(0)}(t)_i=\lim_{n\to\infty}
x^{(n,0)}(t)_i$ exists.}
\vskip2mm

The following proof reproduces the proof of theorem 2.1 in
\cite[p.32]{CMP000} for $d=2$, which applies essentially
unaltered. Define

\be\eqalign{ \d_i(t,n)\defi&|q_i^{(n,0)}(t)-q^{(n+1,0)}_i(t)|,\cr
u_k(t,n)\defi&\max_{q_i\in \L_k}\d_i(t,n),\cr}\Eq{e8.2}\ee
then, for $a=0$, Eq.\equ{e8.1} yields

\be \eqalign{\d_i(t,n)\le&  \int_0^t \frac{\Th}{m}\,d\t\ \big\{
F'_w\d_i(\t,n)\cr
&+F'\sum_j
(\d_j(\t,n)+\d_i(\t,n))\,\big\}\cr}\Eq{e8.3}\ee
where the sum is over the number $\NN_n$ of the particles $q_j(\t)$
that can interact with $q_i(\t)$ at time $\t$; $F'=\max|\partial^2
\f|$ is the maximal gradient of the interparticle force; $F'_w= C
\a(\a+1) \frac{\f_0}{ r^2_\ps}\, n^{\frac{\a+2}\a}$ bounds the
maximum gradient of the walls plus the stirring forces, the bound
follows from Eq.\equ{e6.11}.  The number $\NN_n$ is bounded by theorem
4, Eq.\equ{e6.11}, by $\NN_n\le Cn^{1/2}$ for both $x^{(n,0)}(\t)$ and
$x^{(n+1,0)}(\t)$. Let

\be \h\defi
(1+\frac2\a),\;\; 2^{k_1}\defi 2^{k}+r_n
\Eq{e8.4}\ee
where $r_n$ is the maximum distance a particle can travel in time
 $\le\Th$, bounded by Eq.\equ{e6.11} by $C\, r_\f\,n^{1/2}$. Then

\be \frac{u_k(t,n)}{r_\f}\le \,C\, n^\h \int_0^t
\frac{u_{k_1}(s,n)}{r_\f} \frac{ds}\Th\Eq{e8.5}\ee
($C$ is a function of $\EE$ as agreed in Sec.\ref{sec5}).
Eq.\equ{e8.5} can be iterated $\ell $ times if $2^k+ C\, \ell n^{1/2}
<2^{n}$, {\it i.e.} $\ell=\frac{2^{n}-2^k}{2C n^{1/2}}$ which is
$\ell\,> \,c \, 2^{n/2}\, \d_{k<n}$ for $n$ large.

By
Eq.\equ{e6.11} $u_k(t,n)$ is $\le C\, n^{1/2}$ so that for  $n>k$,

\be\frac{ u_k(n,t)}{r_\f}\le
\, C'\, \frac{(n^{\h})^{\ell+1}}{\ell!} n^{1/2}
\le C\, 2^{- 2^{n/2}c } \Eq{e8.6}
\ee
for suitable $C',C,c>0$ ($n$--independent functions of $\EE$). Hence the
evolutions locally ({\it i.e.} inside the ball $\L_k$) become closer
and closer as the regularization is removed ({\it i.e.}  as
$n\to\infty$) and very fast so.

If $q_i(0)\in\L_k$, for $n>k$ it is

\be q_i^{(0)}(t)=q_i^{(k,0)}(t)+\sum_{n=k}^\infty
(q^{(n+1,0)}_i(t)-q^{(n,0)}(t))
\Eq{e8.7}\ee
showing the existence of the dynamics in the thermodynamic limit
 because also the inequality,  for $n>k$,

\be\frac{|\dot q^{(n,0)}(t)-\dot
 q^{(n+1,0)}(t)|}{v_1}\le \,C\, 2^{- 2^{n/2}c }\Eq{e8.8}\ee
follows from Eq.\equ{e8.6} and from $\dot q^{(n,0)}(t)-\dot
 q^{(n+1,0)}(t)=\int_0^t\big( f_i(q^{(n,0)}(\t))d\t
 -f_i(q^{(n+1,0)}(\t))\big)d\t$. Or, for $n>k$,

\be|x^{(n,0)}_i(t)-x^{(0)}_i(t)|\le \,C\, e^{-c 2^{nd/2}}\Eq{e8.9}\ee
calling $|x_i-x_i'|\defi {|\dot q_i-\dot q'_i|}/{v_1}+
{|q_i-q'_i|}/{r_\f}$.

Hence the proof of the existence of the dynamics in the Hamiltonian
case and in the thermodynamics limit is complete and it
yields concrete bounds as well, {\it i.e}
\vskip2mm

\noindent{\bf Theorem 7:} {\it There are $C(\EE),c(\EE)^{-1}$,
increasing functions of $\EE$, such that the Hamiltonian evolution
satisfies the local dynamics property and if $q_i(0)\in\L_k$

\be \eqalign{
&|\dot q_i^{(n,0)}(t)| \le v_1 C(\EE) \, k^{\frac12} ,\cr
&{\rm distance}( q_i^{(n,0)}(t), \partial(\cup_j \O_j\cap \L))\ge
c(\EE) k^{-\frac1\a} r_\ps\cr
&\NN_i(t,n)\le C(\EE) \, k^{1/2}\cr
&|x^{(n,0)}_i(t) -x^{(0)}_i(t)|
\le C(\EE) r_\f e^{-c(\EE) 2^{nd/2}}\cr}
\Eq{e8.10}\ee
for all $n> k$. The $x^{(0)}(t)$ is the unique solution of the Hamilton
equations satisfying  the first three of Eq.\equ{e8.10}.}
\vskip2mm

The uniqueness follows from Eq.\equ{e8.3} and we skip the details,
\cite{CMS005}.

It would also be possible to show the stronger result that
$x^{(0)}(t)\in \HH_{1/d}$: but for the proof of theorem 1 the theorems
5,6,7 are sufficient, hence the proof of the stronger property is
relegated to theorem 9 in the appendix.

The corresponding proof for the thermostatted evolution will be
somewhat more delicate: and {\it it will be weaker} as it will not hold
under the only assumption that $\EE(x)<\infty$ but it will be
necessary to restrict further the initial data to a subset of the
phase space (which however will still have $\m_0$--probability $1$).

\vskip2mm

\noindent{\it Remark:} An immediate consequence is that the {\it
  entropy production} $\s(x)$, see Eq.\equ{e7.3}, is estimated by a
  constant $s(\EE)$ in the $\L_n$--dynamics for $n$ large: {\it i.e} a
  much better estimate than the growth bounded by a power of $n$, see
  Eq.\equ{e7.4}, implied by Eq.\equ{e6.11}.

\def\SEC{Infinite volume thermostatted dynamics}
\section{Infinite volume thermostatted dynamics}\label{sec9}
\iniz

Eq.\equ{e8.1} will be used to compare the Hamiltonian and the
thermostatted evolutions in $\O_j$ with the same initial data assuming
that the initial data satisfy theorem 5, Sec.\ref{sec7}.  We shall see
that the problem will reduce to obtain a better estimate of the
entropy production, {\it i.e.} better than proportional to $n$, as in
Eq.\equ{e7.4}.

Fixing once and for all $\k>0$ to be smaller than the minimum of the
kinetic energy densities of the initial $x$ in the various thermostats
(which is $x$-independent with $\m_0$--probability $1$), the problem
can be solved by restricting attention to a suitable subset of the set
$\XX_{E}\subset \HH_{1/d}$:

\be \XX_{E}\defi\big\{x\,|\, \EE(x)\le E;\, \min_{t\le \Th\atop j>0}
     \frac{K_{j,n}(S^{(n,1)}_tx)}{\f_0}\ge \k 2^{nd}\big\} \Eq{e9.1} \ee
In this section (and in the corresponding appendix E) the constants
$C,C',\dots,$ $c,c',\ldots$ will be functions of $E$ as stated in
Sec.\ref{sec5}.

Consider the bands of points $\x$ at distance $\r_{\O_0}(\x)$
from the boundary $\partial\O_0$ of $\O_0$ within  $r_\f$ or $2r_\f$

\be\eqalign{ & \L_*\defi \{q: \r_{\O_0}(q)\le r_\phi\},
\cr
&\L_{**}\defi \{q: \r_{\O_0}(q)\le 2r_\phi\}\cr}\Eq{e9.2} \ee
By the result in Sec.\ref{sec8} there are $M$ and $V$ (which depend on
$E$) so that for all $x\in \XX_E$ and with the notations
Eq.\equ{e2.8}, for $n$ large enough:

\be\eqalign{ &\max_{t\le \Th} N_{\L_{**}}(S^{(n,0)}_{t}x)
<M \cr &\max_{t\le \Th} \,V_{\L_{**}}(S^{(n,0)}_{t}x)< V-1
\cr}\Eq{e9.3}\ee
Define for $x$ the stopping times

$$\eqalignno{T_{M,\,V;\,n}&(x)\defi \max\big\{t\le
  \min\{t_{\L_n}(x),\Th\}:\, \forall\, \tau\le t,\cr &
N_{\L_{*}}(S^{(n,1)}_{\tau}x) \le M,\ V_{\L_{*}} (S^{(n,1)}_{\tau}x)
\le V \big\}&\eq{e9.4} \cr} $$
Let $C_\x$ the cube with side $r_\f$ centered at a point $\x$ in
the lattice $r_\f \ZZZ^d$, and using the definitions in
Eq.\equ{e2.8}, let

\be\|x\|_n\defi\max_{\x\in\L_n} \frac{
\max (N_{C_\x}(x),\sqrt{E_{C_\x}(x)})}{g_{\l}(\x/r_\f)}.\Eq{e9.5}\ee
with $\frac12<\l<1$.
Split $\XX_{E}=\AA\cup \BB$ where

\be\eqalign{
\AA\defi&\{x\in \XX_{E}: \max_{t\le T_{M,V;\,n}(x)}
    \|S^{(n,1)}_tx\|_{n} \le (\log n)^{\l}\} \cr \BB\defi&\{x\in
    \XX_{E}: \max_{t\le T_{M,V;\,n}(x)} \|S^{(n,1)}_tx\|_{n} >
    (\log n)^{\l}\}.\cr}
\Eq{e9.6}\ee
Fixed, once and for all, $\g>0$ arbitrarily
\vskip2mm

 \noindent{\bf Theorem 8:} {\it In $d=1,2$ there are
 positive constants $C,C',c$ depending only on $E$ such that
 for all $n$ large enough:
\\
(1) if $x\in \mathcal A$ then
 $T_{M,V;\,n}(x)=\Th$,  $S^{(n,0)}_tx$ and $S^{(n,1)}_tx$ are
 close in the sense that for $q_i(0)\in \L_{(\log n)^\g}$

\be \eqalign{
&|q^{(n,1)}_i(t) -q^{(n,0)}_i(t)|\le \,C\, r_\f\, e^{-(\log n)^\g\,c },\cr
&|\dot q^{(n,1)}_i(t) -\dot q^{(n,0)}_i(t)|\le \,C\,v_1\, e^{-(\log
    n)^\g\,c }.
\cr}
\Eq{e9.7}\ee
(2) the set $\BB$ has $\m_0$--probability bounded  by

\be \mu_0(\BB) \le \,C\,e^{-c (\log n)^{2\l}+C' M^2 V}.\Eq{e9.8}\ee
}
\vskip2mm

\noindent{\it Remark:} Since $\l>1/2$ Eq.\equ{e9.8} will imply (by
Borel--Cantelli's lemma) that, with $\m_0$--probability $1$,
eventually $x$ is in $\AA$ and therefore, as soon as Eq.\equ{e9.7}
will have been proved, the thermodynamic limits of $q^{(n,a)}(t)$ will
coincide for $t\le\Th$, $a=0,1$: concluding the proof of theorem 1 as
well.  \vskip2mm

\subsection{Check of Eq.\equ{e9.8}}

We begin by defining the surface, see it symbolic description in Fig.2 below:

\be\eqalign{
\Si_\t\defi&\big\{ x\in\X_E\,| \,||S^{(n,1)}_{\t} x||_n\ge(\log
n)^\l,\cr 
&\forall \, t< \t,\ ||S^{(n,1)}_{t} x||_n< (\log
n)^\l\cr
&T_{M,V;\,n}(x)\ge\t\big\}\cr}\Eq{e9.9}\ee
Recalling the definition of the existence time $t_{\L_n}(x)$ in
Sec.\ref{sec1}, p.\pageref{t}, item (6), we remark that for $x\in
\X_E$, $t_{\L_n}(x)\ge \Th$ so that $S_\t^{(n,1)}x$, $\t\le\Th$ is well
defined.

Moreover $S_\t^{(n,1)}\Si_\t$ is contained in the {\it surface} $\Si'$
(piecewise smooth) of points $x$ for which $S_{t}^{(n,1)}x$ is well
defined for $t<0$ near $0$ and $ ||S_{t}^{(n,1)}x||$ crosses from
below the value $(\log n)^\l$ at time $t=0$.

With the above geometric considerations (see Fig.2) the set $\Si\defi$
$\cup_{\t\le\Th}S_{-\t}^{(n,1)}x\in\Si_\t\subset \Si'$ and for $x\in
\Si$ we define $\th(x)=\max_{\t\le \Th} \big\{\t\,|\,
S_{-\t}^{(n,1)}x\in\Si_\t\big\}$.  \vskip2mm

\eqfig{200}{86}{%
\ins{61}{75}{$x$}
\ins{-1}{57}{$\Si_\t$}
\ins{6}{39}{$\Si_{\t'}$}
\ins{61}{20}{$S^{(n,1)}_{-\th(x)}x$}
\ins{90}{72}{$\Si'\supset \Si$}
}{fig2}{}

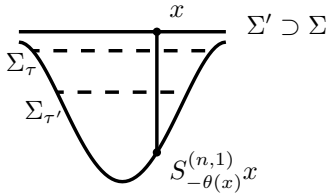
\noindent{Fig.2: \small\it The horizontal ``line'' represents $\Si'$;
  the ``curve'' represents the points $S^{(n,1)}_{-\th(x)}x$, i.e the
  set of points in $\XX_E$ which in time $\th(x)$ reach $\Si'$ and
  determine on it a subset $\Si$; the incomplete (``dashed'') lines
  represent the ``levels'' $\Si_\t, \Si_{\t'}$; the missing parts are
  made of points which are not in ${\cal X}_E$ but have an
  ``ancestor'' in $\mathcal X_E$; the vertical line represents the
  trajectory of the point $S_{-\th(x)}x$.  } \*
The $\Si_\t$ are represented in Fig.2 as dashed lines to remind that
it might be that the trajectories that reach the surface $\Si'$ will
have a value of $\EE(S_{-t}^{(n,1)}x)>E$ or a kinetic energy $<\k
2^{nd}/\f_0$, see Eq.\equ{e9.1}, at some $t\in(0,\th(x))$. 
\*

\0{\it Remark:} We can also say that $\Si$ is the subset of the surface $\Si'$
consisting of the points of $\Si'$ that are reached by trajectories of
points $y\in \X_E$ within a time interval $\le T_{M,V;\,n}(y)$.
\*

In the notations of Appendix D, $\Si$ is
the base and $\th(x)$ the ceiling function.
We then have
 \be
 \mu_0( \mathcal B )\le\int_{\Si'}
 \mu_{0,\Si'}(dy) \int_{0}^{\theta(y)}dt\,w(y)\,
e^{\widehat\sigma(y,t)}
\Eq{e9.10}\ee
where $\mu_{0,\Si'}$ denotes the projection of $\mu_0$ on $\Si'$,
$w=|v_x\cdot n_x|$ and $\widehat\sigma(y,t)\defi$ $
\int_{-t}^0\sigma(y^{(n,1)}(t'))dt'$ is the phase space contraction.
By the definition of $\Si_\t$ it then follows that $T_{M,V;\,n}(x)\ge
\th(x),\, \forall x\in\Si$ and
$|\widehat\sigma(y,t)| \le C M^2 V$.

Writing $k_\xi$ for the smallest integer $\ge (\log n)^\lambda
g_\lambda(\xi/r_\phi)$, $\mu_{0,\Si'}$ almost surely, $\Si'$ splits
into an union over $\xi \in \Lambda_n\cap r_\f\ZZZ^d$ of the union of
$\mathcal S^1_\xi\cup \mathcal S^2_\xi$, where

\be\eqalign{ \mathcal S^1_\xi=&\{y\in \Si':|y\cap C_\xi|=k_\xi, |y\cap
\partial C_\xi|=1\}\cr \mathcal S^2_\xi=&\{y\in \Si':y\cap C_\xi\ni
(q,\dot q), E(q,\dot q)=\wt E_\x\}\cr}\Eq{e9.11} \ee
if $\wt E_\xi\defi \big((\log n)^\lambda g_\lambda(\xi/r_\phi)\big)^2$.

Both   $\mu_{0,\Si'}({\cal S}^1_\x)$ and $\mu_{0,\Si'}({\cal
S}^2_\x)$ are bounded  by

\be \mu_{0,\Si'}({\cal S}^i_\x)\le C e^{C M^2V}\sqrt{n} e^{-c
[(\log n)^\l g_\l(\x/r_\f)]^2 }, \Eq{e9.12}\ee
(for suitable $C,c$, functions of $E$). The proof of Eq.\equ{e9.12}
does not involve dynamics but only classical equilibrium estimates,
{\it the details are expounded in Appendix} E. Summing (as $\l>1$) over
$\x\in\O_j\cap\L_n$ the Eq.\equ{e9.8} follows.

\subsection{Two remarks}

To prove item (1) and Eq.\equ{e9.7}, thus completing the proof of
theorem 8, we shall compare the evolutions $x^{(n,a)}(t)$ with
$a=0,1$, same initial datum $x\in \AA$ and $t\le T_{M,V;\,n}(x)$, the
latter being the stopping time defined in Eq.\equ{e9.4}. We start by
proving that there is $C>0$ so that for all $n$ large enough the
following holds.  \*

\noindent{\bf Lemma 1:} {\it Let $\g>0$. For $x\in\AA$ and $h\ge (\log
  n)^\g$, then
\be\eqalign{
|\dot q^{(n,1)}_i(t)|&\le C\,v_1\, \big(h\,\log n)^\l,
\cr
|q^{(n,1)}_i(t)|&\le r_\f\,(2^h+C \, \big(h\,\log n)^\l).
\cr}
\Eq{e9.13}\ee
for $q_i(0)\in\L_h$ and $t\le\Th$.}
\*

\0{\it Proof:} if $x\in\AA$ and $t\le T_{M,V;\,n}(x)$ then
\be |\dot q^{(n,1)}_i(t)|\le v_1 \big((\log n) \log_+ \frac{
  |q^{(n,1)}_i(t)|+\sqrt2 r_\f}{r_\f} \big)^\l,\Eq{e9.14}\ee
implying: $|q^{(n,1)}_i(t)|\le r(t) r_\f$ if $r(t)\,r_\f$ is an upper
bound to a solution of Eq.\equ{e9.14} with $=$ replacing $\le$ and
initial datum $|q^{(n,1)}_i(0)| \le 2^hr_\f$.  And $r(t)$ can be taken
$r(t)\defi 2^h+ 2v_1 \big((\log n) \log_+ { 2^h}\big)^\l \,
\frac{t}{r_\f}$, for $t\le \Theta$, provided

\be \big((\log n) \log_+{ r(\Theta)+\sqrt2}\big)^\l \le 2
\big((\log n) \log_+{ 2^h}\big)^\l \Eq{e9.15}\ee
which is verified for all $n$ large enough, because $\frac{(h\log
  n)^\l}{2^h}$ vanishes as $n$ diverges (keeping in mind that $h\ge
  (\log n)^\g$).  Thus $|q^{(n,1)}_i(t)|\le \,r_\f\,r(t)$, hence
  $|\dot q^{(n,1)}_i(t)|\le \, r_\f\, C\,\dot r(t)$ for all $t\le
  T_{M,V;\,n}(x)$, {\it i.e.} when Eq.\equ{e9.14} holds:  and the lemma
  is proved. Fix $\g>0$.  \*

\0{\bf Lemma 2:} {\it Let $\NN$ and $\rho$ be the maximal number of
  particles which at any given time $\le\Th$ interact with a particle
  initially in $\L_{k+1}$ and, respectively, the minimal distance of
  any such particle from the walls in either dynamics and for $t\le
  T_{M,V;\,n}(x)$. Then

\be
\NN \le C\,(k \log n)^{\l},\;\; \r \ge\, c\, (k \log n)^{-2\l /\a}
  \Eq{e9.16}\ee
for all integers $k>(\log n)^\g$.}
\*

\0{\it Proof:} As a consequence of lemma 1 and of
theorem 7 the following properties hold for all $n$ large enough and
all $t\le T_{M,V;\,n}(x)$, both for the Hamiltonian and the
thermostatted evolutions.  \vskip2mm

\noindent
(i) for all $q_i\in \L_{k+2}$ and $a=0,1$,
 \be \max_{t\le T_{M,V;\,n}(x)} |q_i^{(n,a)}(t)-q_i| \le C\, r_\f (k
\log n)^{\l},\; \Eq{e9.17}\ee
\\
(ii) particles $\in\L_{k}$ do not interact with those
$\not\in\L_{k+2}$; \vskip2mm

By Eq.\equ{e9.17} we see that if $q_i\in\L_{k+1}$ then
$q_i^{(n,a)}(t)\in \L_{k+2}$ so that, by the definition of the set
$\AA$ and by Eq.\equ{e9.6} (or by theorem 7 in the Hamiltonian case,
recalling that $\l>1/2$), Eq.\equ{e9.16} follows.

\subsection{Check of Eq.\equ{e9.7} and 
comparison of Hamiltonian versus thermostatted motions}

We have now all the ingredients to bound $\d_i(t,n) \defi$ $
|q_i^{(n,1)}(t)- q_i^{(n,0)}(t)|$.  Let $f_i$ be the acceleration of
the particle $i$ due to the other particles and to the walls.  By
Eq.\equ{e9.16} if $q_i\in \L_{k+1}$, $|f_i|\le C\, (k\log n)^{\eta'}$,
$\eta'\defi 2d\,\l\,(1+\frac 1\alpha)$ so that, subtracting the
Eq.\equ{e8.1} for the two evolutions, it follows that for any $q_i\in
\L_{k+1}$ (possibly close to the origin hence very far from the
boundary of $\L_k$ if $n$ is large, because $k>(\log n)^\g$)

$$\eqalignno{ &\d_i(t,n)\le  C \, (k\log n)^{\eta'} 2^{-nd} &\eq{e9.18}
\cr& + \Th
\int_0^t |f_i(q^{(n,1)}(\t))-f_i(q^{(n,0)}(\t))|\,d\t.\cr
}$$
because, recalling Eq.\equ{e9.1}, $|\a_j|$ is bounded
proportionally to $2^{-nd}$.

Let $\ell$ be a non-negative integer, $k_\ell$ such that
\be 2^{k_\ell}=2^k +\ell \,C\, (k \log n)^{\l} \Eq{e9.19}\ee
(see Eq.\equ{e9.17}) and $u_{k_\ell}(t,n)$ the max of
$\d_i(t,n)$ over $|q_i|\le 2^{k_\ell}$.  Then by Eq.\equ{e9.18} and
Eq.\equ{e9.16} and writing $\eta''\defi 2\,d\,\l(1+\frac 2\alpha)$,

$$\eqalignno{ &\frac{u_{k_\ell}(t,n)}{r_\f}\le \,C\, (k\log
  n)^{\eta'} 2^{-nd} 
&\eq{e9.20} 
\cr& +C  (k\log n)^{\eta''}
\int_0^t\frac{ u_{k_{\ell+1}}(s)}{r_\f}\frac{ ds}\Theta.\cr }$$
for $\ell \le \ell^*=2^k/((k\log n)^\l C)$, the latter being the
largest $\ell$ such that $2^{k_\ell}\le 2^{k+1}$.  By Eq.\equ{e9.20}
and Eq.\equ{e9.16}

$$\eqalignno{ & u_{k}(t,n) \le e^{C\, (k\,\log n)^{\eta''}} C (k\log
  n)^{\eta'} 2^{-dn} 
&\eq{e9.21}
\cr& + \frac{(C\, (k\,\log
    n)^{\eta''})^{\ell^*}} {\ell^*!}\; C\,(k \log n)^{\l}.\cr }$$

Thus $u_{k}(t,n)$ is bounded by the r.h.s.\ of the first of
Eq.\equ{e9.7}; analogous argument shows that also the velocity
differences are bounded as in Eq.\equ{e9.7} which is thus proved for
all $t\le T_{M,V;\,n}(x)$.  

On the other hand given $q_i(0)$ with $|q_i(0)|/r_\f\le 2^{k_0}$ it
is, for $n> e^{k_0^{1/\g}}$ and $i$ fixed,
$|q_i^{(n,1)}(t)-q^{(n,0)}_i(t)|<u_{(\log n)^\g}(t,n)\le C 2^{dn/2}$,
{\it i.e.} for $n$ large $q^{(n,1)}_i(t)$ is closer
than $r_\f$ to $q^{(n,0)}_i(t)$. Hence, remarking that we know
``everything'' about the Hamiltonian motion we can use such knowledge
by applying Eq.\equ{e9.7} to particles which are initially within a
distance $r_\f 2^{k_0}$ of the origin, with $k_0$ fixed arbitrarily,

Therefore the number of particles in $q^{(n,1)}_i(t)$ which are in
$\L_*$ is smaller than the number of particles of $q^{(n,0)}_i(t)$ in
$ \L_{**}$ which is bounded by $M$.  An analogous argument for the
velocities allows to conclude that Eq.\equ{e9.3} hold in $\L_{**}$
also for the thermostatted motion ($a=1$, being valid for the
Hamiltonian motion in the smaller $\L_*$, given the closenes of the
positions and speeds), $T_{M,V;\,n}\equiv \Th$ with $\m_0$--probability
$1$.  

Applying again Eq.\equ{e9.7} the proof of theorem 8 is complete:
with $\g=2$ (but any $\g>0$ would also lead to a corresponding
result).

\def\SEC{Concluding remarks}
\section{Concluding remarks}\label{sec10}
\iniz

Equivalence between different thermostats is widely studied in the
literature and the basic ideas, extended here, were laid down in
\cite{ES93}. A clear understanding of the problem was already set up
in comparing isokinetic, isoenergetic and Nos\'e-Hoover bulk
thermostats in \cite{ES93}, where a history of the earlier results is
presented as well, see also \cite{Ru00,Ga008d}.

There are, since a long time, studies of systems with free
thermostats, starting with \cite{FV63}. Such thermostats are somewhat
pathological and may not always lead to the stationary states that
would be expected: as exemplified in the case of simple spin chains,
\cite{ABGM72, Le71}.  More recently similar or identical thermostat
models built with free systems have been considered starting with
\cite{EPR99}.

The case of dimension $3$ is very similar: it is not difficult to
prove, that the key bounds \equ{e6.9} hold; however a naive
application of the ideas developed in \cite{CMP000} to prove that
$R_n(t)$ satisfies Eq.\equ{e6.11} is not possible.

Isokinetic thermostats should be treated in a very similar way,
\cite{Ga008d}: the extra difficulty is that the entropy production in
a finite time interval receives a contribution also from the time
derivative of the total energy of the reservoirs, \cite{Ga008d}, and
further work seems needed.

 More general cases, like Lennard-Jones interparticle potentials are
more difficult, see \hbox{\cite{BGGZ05}}.  Finally here the
interaction potential has been assumed smooth: singularities like hard
core could be also considered at a heuristic level. It seems that in
presence of hard cores plus smooth repulsive potentials all estimates
of Sec.\ref{sec5},\ref{sec6} are still valid but the existence of the
limiting motion as $\L\to\infty$ remains a difficult point because of
the discontinuities in the velocities due to collisions.

\def\SEC{Appendices}
\section{Appendices}\label{sec11}
\iniz

\subsection{Appendix: Sets of full measures}

There are $c_0$ and $R_0$ and a strictly positive, non decreasing
function $\g(c)$, $c\ge c_0$, so that $\forall c\ge c_0,\forall R\ge
R_0$,

\be \mu_0\Big( W(x,0,R) \ge C R^d\Big) \le e^{-\g(c) R^d} \Eq{e11.1}
\ee
If $g:\ZZZ^d \to \RRR_+$, $g(i)\ge 1$, $c\ge c_0$, the probability

\be \mu_0\Big( \cap _{i\in \ZZZ^d, r \ge g(i)}W(x; i, r) \le
  c\,r^d \Big) \Eq{e11.2} \ee
is $\ge 1- \sum_{i\in \ZZZ^d, r \ge g(i)} e^{-\g(c) r^d}$ with  the sum
being bounded proportionally to the sum $ \sum_{i\in \ZZZ^d} e^{-\g(c) [g(i)]^d}
$ which converges if $g(i) \ge c'(\log_+|i|)^{1/d}$, with $c'$
large enough.

\subsection{{\bf Appendix: Choice of $R_n(t)$}}

The proof of the inequalities Eq.\equ{e6.4},\equ{e6.9} yields 
$\forall t\le \Theta$ that $ W(S^{(n)}_t x,R) \le c W(x ,R +\int_0^t V_n) $
provided $R$ is such that $ \frac{R +\int_0^t V_n}{R} \le 2$, which is
implied by $ R\ge R_0 + \int_0^t V_n(s)ds,\ R_0 \ge 0 $.  The maximal
speed $V(t)$ at time $t$ is bounded by $V(t) \le v_1\sqrt {2
W_n(S^{(n)}_t x,R)}$.  Choosing $R_0=R_n(0)=n^{\frac1d}$ we get $V_n(t)
\le C' v_1 R_n(t)^{d/2}\le C v_1 n^{\frac12}$: such choice is the weakest that
still insures that the set of initial data has $W(x,0,R)/R^d$ finite
with $\m_0$--probability $1$, see appendix A.

\subsection{{\bf Appendix: The Hamiltonian motion is a flow in
$\HH_\z$, $\z\ge\frac1d$}}


The following theorem is obtatined by a straightforward adaptation to
the case $d=1,2$ of theorem 2.2 in \cite{CMS005}.  \vskip2mm

\noindent{\bf Theorem 9:}  {\it Let $d=1,2$, $\z\ge 1/d$, $E>0$. Then,
  given any $\Th$ there is $E'$ (depending on $\z,\Th,E$)
so that for all $x$ such that $\EE_\z(x)\le E$

\be \EE_\z( S^{(0)}_tx )\le E',\qquad \hbox{for all}\
  t\le\Th\Eq{e11.3}\ee
so that the evolution $x\to S_t^{(0)}x$ is a flow in all spaces
$\HH_\z$, $\z\ge 1/d$.}

\vskip2mm So far, for the sake of definiteness, $\z=1/d$ has been
assumed: therefore in the following proof the quantity $\z$ has to be
intended equal to $1/d$; however $\z$ is left undetermined
because the proof would still hold for arbitrary $\z\ge1/d$, if larger
values were consistently assigned to it since the beginning of this
paper, under the {\it essential restriction $d\le2$}.
\vskip2mm

\noindent{\bf Proof.} Let $x^{(n,0)}_t\defi S^{(n,0)}_t x$ and consider

\be \widetilde W\big(x^{(0)}_t, \x,\r \big),\quad {\rm for}\quad
\r \ge (\log_+
({|\x|}/{r_\f}))^{\z} \Eq{e11.4} \ee
with $\wt W$ defined as in Eq.\equ{e5.7} with no restriction in the
sums over $q,q'$. 
Let $n_\x-1$ be the smallest integer such that $\Lambda_{n_\x-1}$ contains
the ball of center $\x$ and radius $\r\, r_\f$.  Then $\forall t\le \Th$

\be\eqalign{ \widetilde W&\big(x^{(0)}_t, \x, \r \big)\le \widetilde
W_{n_\x}\big(x^{(n_\x,0)}_t, \xi,\r \big)\cr &+ |\widetilde
W\big(x^{(0)}_t, \x,\r \big)- \widetilde W_{n_\x}\big(x^{(n_\x,0)}_t, \x,\r
\big)| \cr} \Eq{e11.5} \ee
The motions $x^{(n_\x,0)}_t$ and $x^{(0)}_t$ are very close for all
points which initially are in $\L_{n_\x-1}$: by
Eq.\equ{e8.6}--\equ{e8.8} the difference of positions and velocities
are bounded by $C \exp{-c\, 2^{n_\x/2}}$.

Setting $\ch_\x(q)$ equal to the smoothed characteristic function
$\ch_\x(q,\r)$ introduced in Eq.\equ{e5.7}, $\chi_\x(q_i^{(n,0)}(t))$
and $\chi_\x(q_i^{(0)}(t))$ force their arguments to be within
$\L_{k_1}$ if $2^{k_1}=2^{n_\x-1} + \r\ll 2^{n_\x}$. Hence the
inequality

\be\eqalign{
&|\widetilde W\big(x^{(0)}_t, \x,\r \big)- \widetilde W_{n_\x}
\big(x^{(n_\x,0)}_t,
 \xi,\r \big)|\cr
&\le  C' e^{- c n_\x}\sum \chi_\x(q_i^{(n_\x,0)}(t))\{|{\dot
 q_i}^{(n_\x,0)}(t)|+ \NN_i(t)\} \cr
 &\le   C\,e^{- c n_\x} \sup_{\x'
 \in \L_{n_\x}}\widetilde W\big(x^{(n_\x,0)}_t,\x', \r +\r_{n_\x}\big)\cr}
\Eq{e11.6}
\ee
where $\NN_i=$ number of points in $x^{(n_\x,0)}_t$ which
interact with $q_i^{(n_\x,0)}(t)$ and $\r_{n_\x}\defi$$ \int_0^\Th
V_{n_\x}(\t)d\t$: recall that $V_{n_\x}(t)= C\,v_1\, n_{\x}^{\z d/2}$, see
Eq.\equ{e6.10} (using the better estimate Eq.\equ{e6.11}, $V_n\le C\,v_1\,
n^{\frac12}$, valid for all $\z$, would lead to the same end result
because $\z d\ge 1$).  Actually $e^{- c n_\x}$ could be replaced by
$\exp{- (c\, 2^{n_\x/2})}$ as in Eq.\equ{e8.6}--\equ{e8.8}.
\vskip2mm

Consider first the case of $\r$ large, say $\r> \r_{n_\x}= O(n_\x^\z)$.
Then $\widetilde W\big(x^{(0)}_t; \x,\r \big)$ can be estimated by
remarking that the argument leading to Eq.\equ{e6.9} remains unchanged
if $R(t)=\r+\int_0^t V_{n_\x}(\t)d\t$ and $R(t,s)=R(t)+\int_s^t
V_{n_\x}(\t)d\t$ are used instead of the corresponding
$R_{n_\x}(t),R_{n_\x}(t,s)$ (as long as $\r\ge0$).
Then

\be \widetilde W(x^{(0)}_t; \x,\r+ \r_{n_\x}\big)\le C \widetilde W(x,
\r+ 2 \r_{n_\x} \big) \Eq{e11.7} \ee
as in the first of Eq.\equ{e6.8}.

Suppose $\r_0-\r_{n_\x}> g_\z(\x/r_\f)$, {\it i.e.} if $\r_0> C
n_\x^\z$, then $\widetilde W(x^{(0)}_t; \x,\r_0)\big)\le C'
\widetilde W(x,\r_0 +\r_{n_\x})\big)\le C'' (\r_0+\r_{n_\x}))^d\le C
\r_0^d$: hence only the values of $(n_\x-1)^\z\le \r_0\le C n_\x^\z$
are still to be examined.

In this case, however, the bound $\widetilde W(x^{(0)}_t; \x,\r+
\r_{n_\x}\big)\le C \widetilde W(x, \r+ 2 \r_{n_\x} \big)$ involves
quantities $\r, \r_{n_\x}$ with ratios bounded above and below by a
constant, hence $\widetilde W(x^{(0)}_t; \x,\r) $ is bounded by
$\widetilde W(x; \x,\r+ Cn^{\z})\le C' \r^d $.

Conclusion: there is $C>0$, depending only on $\EE_\z$ and for all
$\r> g_\z(\x/r_\f), \,t\le\Th$ it is $W(x^{(0)}_t; \x,\r)\le C\, \r^d
$.

\subsection{Appendix: Quasi invariance}

A probability distribution $\m$ on a piecewise regular manifold $M$ is
{\it quasi invariant} for a flow $x\to S_t x$ generated by a
differential equation $\dot x= v_x$ if $e^{-\l(t)}\le
\m(S_{-t} dx)/\m(dx)\le e^{\l(t)}$ and $\l(t)<\infty$.

Suppose given $\Th>0$, a piecewise smooth surface $\Si\subset M$ with
unit normal vector $n_x$ and a
``stopping time'' $x\to\th(x)\le \Th$ defined on $\Si$ consider all points
$x\in\Si$ which are reached {\it for the first time} in positive time
$t\le \th(x)$ from data $y\not\in\Si$. Call $E$, the set of such
points, the {\it tube with base $\Si$ and ceiling $\th(x)$}.

The probability distribution $\m$ is {\it quasi invariant} with
respect to $\Si$ and to the stopping time $x\to\th(x)$ if it is
absolutely continuous with respect to the volume measure, its density
$r(x)$ is continuous and $e^{-\l}\le \m(S_{-t}dx)/\m(dx)\le e^{\l}$
for some $\l>0$ and for all $0\le t\le \th(x)$: this is referred to by
saying that $\m$ is quasi invariant with respect to the stopping time
$\th(x)$ on $\Si$: symbolically $\m$ is $(\Si,\th(x))$--$\l$-quasi
invariant.

Then the following {\it Sinai's lemma}, \cite{Si974,MPP975,FD977}, holds:
\vskip2mm

\noindent{\bf Lemma:} {\it If $\m$ is $(\Si,\th(x))$--$\l$-quasi invariant the
  integral of any non negative function $f$ over the tube with base
  $\Si$ and ceiling $\th(x)$ can be bounded  by

$$\eqalignno{
\int_E &f(y)\m(dy) \le e^\l \int_\Si \int_0^{\th(x)}  r(x)\,f(S_{-\t}x)
\,v_x\cdot n_x\, ds_xd\t,\cr
&
\ge e^{-\l} \int_\Si \int_0^{\th(x)} r(x)\, f(S_{-\t}x)
\,v_x\cdot n_x\, ds_xd\t&{\rm\eq{e11.8}}\cr}$$
}
\vskip2mm

The lemma can been used to reduce dynamical estimates to equilibrium
estimates.
\vskip2mm

\noindent{Proof:} Let the trajectory of a point $y$ which reaches
$\Si$ within the stopping time at $x\in\Si$ be parameterized by the
time $\t$ and let $ds_x$ be the surface element on $\Si$. Then the set
of points into which the parallelepiped $\D$ with base $ds_x$ and
height $d\t$ becomes a parallelepiped $S_{-t}\D$ with base $S_t ds_x$
and the same height $d\t$. Therefore the measure of $\m(S_t\D)$ is
$e^{-\l} \le \frac{\m(S_t\D)}{\m(\D)} \le e^\l$ hence the integral of
any positive function $f(y)$ over the set $E$ can be bounded above and
below by the integral of $\int_{\Si}\int _0^{\th(x)} f( S_t x) \r(x)
ds_x d\t$ if $\r(x)ds_x d\t$ is the measure of $\D$: the latter is
$r(x) v_x\cdot n_x ds_xd\t$. Therefore $\r(x)=r(x)\,v_x\cdot n_x$.

\subsection{Appendix: Proof of Eq.\equ{e9.12}}

The factor $e^{-C' M^2V}$ arises because of the entropy bound ({\it
  i.e.} from the phase space contraction estimate within the stopping
time).  Therefore it is sufficient to find a bound to the integral in
Eq.\equ{e9.10} {\it without the factor $e^{\widehat\s(y,t)}$}.

Consider first the case of ${\cal S}^1_\x$. By Eq.\equ{e9.11} if $y\in
\mathcal S^1_\x$ then $|y\cap C_\xi|=k_\xi$ and there is $(q,\dot
q)\in y$ with $q\in \partial C_\xi$. 

Remark that $y$ is the configuration reached starting from an initial
data $x\in \X_E$ within a time $<T_{M,V;\,n}(x)<t_{\L_n}(x)$: hence
Eq.\equ{e6.11} applies. By Eq.\equ{e6.11} $w(y)\le |\dot
q|\le v_1\,C\,\sqrt{n}$ so that

\be\eqalign{
&\int_{\mathcal S^1_\xi}  \mu_{0,\Si'}(dy)\int_{0}^{\theta(y)}dt\,w(y)\,
\cr
&\le \Theta v_1 C\sqrt{n}\int \mu(dx) \frac
{J_1}{Z_{C_\xi}(x)}\cr}\Eq{e11.9} \ee
where  $\mu(dx)$ is the  $\mu_0$-distribution
of  configurations $x$ outside $C_\xi$ and

\be J_1= \int_{\partial C_\xi} dq_1 \int_{C_\xi^{k_\xi-1}}
\frac{dq_2\dots dq_{k_\xi}}{(k_\xi-1)!}  \int_{\rrr^{dk_\xi}} d\dot q
e^{-\beta_j H(q,\dot q|x)}\Eq{e11.10}\ee
The estimate of the {\it r.h.s.} of Eq.\equ{e11.9}, as remarked, is an
``equilibrium estimate''.  By superstability, \cite{Ru970}, the
configurational energy $U(q|x)$ $\ge bk_\xi^2 -b'k_\xi$, so that $J_1$
is bounded by:

 \be B\,e^{-\beta_j(bk_\xi^2 -b'k_\xi)} \frac{|C_\xi|^{k_\xi-1}|
\partial C_\xi|}{(k_\xi-1)!} (\frac{2\pi } {\beta_jm})^{\frac{d}{2}
{k_\xi}}\Eq{e11.11}\ee
while $\int \mu(dx) \frac 1{Z_{C_\xi}(x)}\le 1$ because
$Z_{C_\xi}(x)\ge 1$: and the bound can be summed over $k_\x$.  Thus
the contribution from ${\cal S}_1$ to Eq.\equ{e11.9} is bounded by

\be C' e^{C M^2 V }\sqrt{n}\, e^{-b [(\log n)^\l g_\l( \x
     /r_\f)]^{2}}\Eq{e11.12}\ee
with $C,b$ suitable positive constants.  Since $\l>1/2$, this is
summable over $\x$ and yields the part of the Eq.\equ{e9.8} coming
from the integration over ${\mathcal S^1_\x}$.

Let, next, $y\in \mathcal S^2_\xi$ and let $(q,\dot q)$ as in
\equ{e9.11}. The function $w$ is

\be w = \frac{|d E(q,\dot q)/dt|}{|{\rm
 grad}E(q,\dot q)|} \Eq{e11.13}\ee
$|d E(q,\dot q)/dt| \le C\,|\dot q| n^{1/2}$ because $d E/dt$ is the
work on the particle $(q,\dot q)$ done by the pair interactions
(excluding the wall forces). It is then bounded proportionally to the
number of particles which can interact with $(q,\dot q)$, which, by
theorem 4, is bounded proportionally to $n^{1/2}$ (as the total
configuration is in $\Si'$). On the other hand, $|{\rm grad}E(q,\dot
q)| = \sqrt{|\partial \psi(q)|^2 + m|\dot q|^2}\ge \sqrt{m}|\dot q|$
hence $w\le C n^{1/2}$ again by Eq.\equ{e6.11} and the remark
preceding Eq.\equ{e11.9}.

Then, analogously to \equ{e11.9}, the integral under consideration is
bounded by $C e^{C' M^2 V }\sqrt{n}$ ($C,C'$ are suitable constants
functions of $E$) times an equilibrium integral $\int \mu(dx) \frac
{J_2}{Z_{C_\xi}(x)}$
with $J_2$ defined by:

$$\eqalignno{ \sum_k & \int_{C_\xi^{k-1}\times \rrr^{k-1}}
  \frac{dq_2\dots dq_k d\dot q_2\ldots d\dot q_{k}}{(k-1)!}e^{-\beta_j
    K(\dot q_2,..,\dot q_k)}\cr & \cdot e^{-\beta_j \tilde E_\xi}
  \,{\rm area}(\{E(q,\dot q)=\tilde E_\x\})&\eq{e11.14} \cr} $$
\vskip2mm

\noindent{}where the ${\rm area} (\{E(q,\dot q)=\tilde E_\x\})$ is the
area of the surface $\{(q,\dot q):E(q,\dot q)=\tilde E_\xi\}$ in
$\RRR^{2d}$ (the $\wt E_\x$ is defined in \equ{e9.11}).  Then $J_2$ is
bounded by

\be\sum_k \frac{B}{(k-1)!} \Big(|C_\xi| \big(\frac{2\pi }
       {\beta_jm}\big)^{\frac{d}2}\Big)^{(k-1)} |C_\xi| (\tilde
       E_\xi)^{(d-1)/2} e^{-\beta_j \tilde E_\xi} \Eq{e11.15}\ee
so that, suitably redefining $C,C'$ (functions of $E$), the
contribution from ${\cal S}_2$ is bounded by 

\be C' e^{C M^2 V }\sqrt{n} e^{-\frac{\beta_j}{2} [(\log n)^\l g_\l(
\x /r_\f)]^{2}} \Eq{e11.16} \ee
and Eq.\equ{e9.8} follows from Eq.\equ{e11.12} and \equ{e11.16}.
\*

\subsection{Appendix: Regularized thermostatted dynamics}

Consider $N$ particles in $\cup_{j}\O_j\cap\L$ with a configuration of
immobile particles outside $\L$. The analysis in \cite{MPPP976} can be
followed and the solution of the equations of motion can be defined on
the set $\G^+$ consisting of the configurations $x$ in which $1$ of
the particles is at $\x$ on the boundary $\partial\L$, where elastic
collisions take place, with normal speed $\dot q\cdot n(\x)>0$. The
time evolution makes sense until the time $\t_+(x)$ of next collision;
it can then be continued after the elastic collision because, apart
from a set of zero volume, the normal speed of the collision can be
assumed $\ne0$, until the time $t^*(x)>0$, {\it if any}, in which the
total kinetic energy in one of the containers $\O_j$, $j\ge0$,
vanishes.

Remark that even in the cases in which the kinetic energy
$\k_{\min}(x)\defi\min_{j>0}K_{j,\L}(S^{(\L,1)}_t x)$ vanishes as
$t\to t^*(x)<+\infty$, the limit as $t\to t^*(x)$ of
$K_{j,\L}(S^{(\L,1)}_t x)$ and of all speeds and positions will exist
(because the accelerations $\a_j\dot q_{ji}$ are bounded by
$\max|\partial \f| N^2$ uniformly in $\k_{\min}$, using Schwartz'
inequality).

Hence a map $T$ between $\G^+$ into itself, mapping one collision to
the outcome of the next, is defined for almost all points of $\G^+$,
\cite{MPPP976}, unless the point $x$ evolves into one with
$\k_{\min}(x)=0$.

Restricting attention to the points of $x\in \G^+$ whose kinetic
energies in any thermostat do not vanish for $0<t\le\t_+(T^px)$, as in
\cite{MPPP976}, for $p\le p^*$ the dynamics $S^{(\L,1)}_tx$ is well
defined up to the time $\th^*(x)=\sum_p^{p^*} \t^+(T^px)$. The value
of $p^*$ is $p^*=+\infty$ unless for some value $p^*$ the particles of
$T^{p^*}x$ grind to a halt before the next collision (which would,
therefore, remain undefined since the equations of motion become
signular). In the latter case a time $t_\L(x)$ is defined signaling
the moment in which the singularity occurs (an event not considered in
the quoted reference because in the Hamiltonian equations considered
there was no singularity of this kind).

Until the time $t_\L(x)\ge \th^*(x)$ the dynamics exists and the only
question relevant for us is whether $\th^*(x)<\min(t_\L(x),\Th)$: this
would mean that there are infinitely many collisions with the walls
and $[0,\th^*(x)]$ would become the natural time of existence of the
evolution rather than the smallest between $t_\L(x)$ and $\Th$, as
used in this paper.

Suppose that $\th^*(x)<\Th$ and that $K_{j,\L}(S^{(\L,1)}_t
 x)>\frac1h$ for $t<\th^*(x)$: call $\G^+_h$ such points $x$. Then the
 volume contraction of the distribution $\lis\m_0$ obtained by
 conditioning $\m_0$ to the particles outside $\L$ is bounded
 uniformly in all subintervals of $[0,\th^*(x))$: hence a set $\D$ in
 $\G^+_h$ generates a ``tube'' $\D^*=\cup_{x\in\D} \cup_{0\le t\le
 t_+(x)} S^{(\L,1)}_tx$ and the volume $\lis\m_0(T^n\D^*)\ge
 \l\,\lis\m_0(\D^*)$ where $\l$ is a lower bound on the
 $\lis\m_0$-volume contraction over any time interval in $[0,\Th]$
 over which the dynamics is defined.  The quantity $\l$ is bounded for
 all $n$ such that $\sum_{p=0}^n \t^+(T^px)<\th^*(x)$.

Therefore the sets $T^k \D^*$ cannot be disjoint for all $k$ unless
 $\sum_k \t^+(T^kx)\ge t_\L(x)$: {\it i.e.} this remark takes the
 place of Poincar\'e's recurrence theorem used in \cite{MPPP976} and
 allows us to conclude that until the thermostats kinetic energies are
 all positive the regularized dynamics exists and the elastic
 collisions with the boundary of $\L$ cannot accumulate in time.

This means that the evolution proceeds until the first time $t_\L(x)$
 (if any) when some of the $K_{\L,j}$ vanishes.

\noindent{\bf Acknowledgements:} This work has been partially suppor\-ted
also by Rutgers University.

\bibliography{0Bib}

\begin{thebibliography}{10}

\bibitem{Ga008d}
G.~Gallavotti.
\newblock On thermostats: {I}sokinetic or {H}amiltonian? finite or infinite?
\newblock {\em Chaos}, 19:013101 (+7), 2008.

\bibitem{FV63}
R.P. Feynman and F.L. Vernon.
\newblock The theory of a general quantum system interacting with a linear
  dissipative system.
\newblock {\em Annals of Physics}, 24:118--173, 1963.

\bibitem{Ru970}
D.~Ruelle.
\newblock Superstable interactions in classical statistical mechanics.
\newblock {\em Communications in Mathematical Physics}, 18:127--159, 1970.

\bibitem{Ga06c}
G.~Gallavotti.
\newblock Entropy, thermostats and chaotic hypothesis.
\newblock {\em Chaos}, 16:043114 (+6), 2006.

\bibitem{Ga00}
G.~Gallavotti.
\newblock {\em Statistical Mechanics. A short treatise}.
\newblock Springer Verlag, Berlin, 2000.

\bibitem{Ga008c}
G.~Gallavotti.
\newblock Thermostats, chaos and {O}nsager reciprocity.
\newblock {\em Journal of Statistical Physics}, 134:1121--1131, 2009.

\bibitem{MPPP976}
C.~Marchioro, A.~Pellegrinotti, E.~Presutti, and M.~Pulvirenti.
\newblock On the dynamics of particles in a bounded region: A measure
  theoretical approach.
\newblock {\em Journal of Mathematical Physics}, 17:647--652, 1976.

\bibitem{Ru99}
D.~Ruelle.
\newblock Smooth dynamics and new theoretical ideas in non-equilibrium
  statistical mechanics.
\newblock {\em Journal of Statistical Physics}, 95:393--468, 1999.

\bibitem{Ru01}
D.~Ruelle.
\newblock Entropy production in quantum spin systems.
\newblock {\em Communications in Mathematical Physics}, 224:3--16, 2001.

\bibitem{CMP000}
E.~Caglioti, C.~Marchioro, and M.~Pulvirenti.
\newblock Non-equilibrium dynamics of three-dimensional infinite particle
  systems.
\newblock {\em Communications in Mathematical Physics}, 215:25--43, 2000.

\bibitem{FD977}
J.~Fritz and R.L. Dobrushin.
\newblock Non-equilibrium dynamics of two-dimensional infinite particle systems
  with a singular interaction.
\newblock {\em Communications in Mathematical Physics}, 57:67--81, 1977.

\bibitem{Si974}
{Ya.}~G. Sinai.
\newblock The construction of the cluster dynamics of dynamical systems in
  statistical mechanics.
\newblock {\em Moscow University Mathematics Bulletin}, 29:124--129, 1974.

\bibitem{MPP975}
C.~Marchioro, A.~Pellegrinotti, and E.~Presutti.
\newblock Existence of time evolution for $\nu$ dimensional statistical
  mechanics.
\newblock {\em Communications in Mathematical Physics}, 40:175--185, 1975.

\bibitem{LR969}
O.~Lanford and D.~Ruelle.
\newblock Observables at infinity and states with short range correlations in
  statistical mechanics.
\newblock {\em Communications in Mathematical Physics}, 13:194--215, 1969.

\bibitem{CMS005}
G.~Cavallaro, C.~Marchioro, and C.~Spitoni.
\newblock Dynamics of infinitely many particles mutually interacting in three
  dimensions via a bounded superstable long-range potential.
\newblock {\em Journal of Statistical Physics}, 120:367--416, 2005.

\bibitem{ES93}
D.~J. Evans and S.~Sarman.
\newblock Equivalence of thermostatted nonlinear responses.
\newblock {\em Physical Review E}, 48:65--70, 1993.

\bibitem{Ru00}
D.~Ruelle.
\newblock A remark on the equivalence of isokinetic and isoenergetic
  thermostats in the thermodynamic limit.
\newblock {\em Journal of Statistical Physics}, 100:757--763, 2000.

\bibitem{ABGM72}
D.~Abraham, E.~Baruch, G.~Gallavotti, and A.~Martin-L{\"o}f.
\newblock Dynamics of a local perturbation in the {$X-Y$} model ({II}).
\newblock {\em Studies in Applied Mathematics}, 51:211--218, 1972.

\bibitem{Le71}
J.~L. Lebowitz.
\newblock {\em Hamiltonian flows and rigorous results in nonequilibrium
  statistical mechanics}, volume Ed. S.A. Rice, K.F.Freed, J.C.Light of {\em
  Proceedings of the VI IUPAP Conference on Statistical Mechanics}.
\newblock University of Chicago Press, Chicago, 1971.

\bibitem{EPR99}
J.~P. Eckmann, C.~A. Pillet, and L.~Rey Bellet.
\newblock Non-equilibrium statistical mechanics of anharmonic chains coupled to
  two heat baths at different temperatures.
\newblock {\em Communications in Mathematical Physics}, 201:657--697, 1999.

\bibitem{BGGZ05}
F.~Bonetto, G.~Gallavotti, A.~Giuliani, and F.~Zamponi.
\newblock Chaotic {H}ypothesis, {F}luctuation {T}heorem and {S}ingularities.
\newblock {\em Journal of Statistical Physics}, 123:39--54, 2006.

\end{thebibliography}
\small \bibliographystyle{unsrt}
\def\SEC{\small References}
\vskip3mm

\noindent{e-mails: \\
\tt giovanni.gallavotti@roma1.infn.it, \\
 presutti@mat.uniroma2.it}

\hfill Roma, \today
\end{document}